\documentclass[
aps,
pra,
reprint,
superscriptaddress,
amsmath,
amssymb
]{revtex4-2}

\usepackage{graphicx}
\usepackage{hyperref}
\usepackage{physics}
\usepackage{comment}

\begin{document}

\title{High-Harmonicity Planar Penning Traps for Single-Electron Qubits}

\author{Kedar Mal}
\email{kedarmaljangirphy@gmail.com}
\affiliation{Indian Institute of Technology Delhi, New Delhi, 110016, India}
\affiliation{Inter-University Accelerator Centre, New Delhi 110067, India}

\author{A.N. Agnihotri}
\affiliation{Indian Institute of Technology Delhi, New Delhi, 110016, India}

\author{Sugam Kumar}
\affiliation{Inter-University Accelerator Centre, New Delhi 110067, India}

\author{Wolfgang Quint}
\affiliation{GSI Helmholtzzentrum f\"ur Schwerionenforschung, 64291 Darmstadt, Germany}

\author{Manuel Vogel}
\affiliation{GSI Helmholtzzentrum f\"ur Schwerionenforschung, 64291 Darmstadt, Germany}

\date{\today}

\begin{abstract}
We present a detailed account of the design choices required for planar Penning traps that feature a highly harmonic confining potential. High harmonicity is indispensable for a number of applications, particularly for confinement of single electrons as qubits in quantum-information processing. The present work extends previous studies [J. Goldman and G. Gabrielse, Phys. Rev. A \textbf{81}, 052335 (2010)] by a fully analytic treatment of finite electrode gaps and their relevance for small traps on the millimeter size scale and below, when relative gap sizes are non-negligible. We derive the overall trap potential with a particular focus on different models of finite-gap potentials and show how to find the optimum trap geometry and electrode voltages to minimize anharmonicities. The analytic calculations are compared with detailed finite-element simulations.
\end{abstract}

\maketitle

\section{Introduction}
\label{one}
Planar Penning traps are an essentially two-dimensional variation on the well-established Penning trap concept in three dimensions \cite{book}. They have been brought forward in the framework of quantum-information processing (QIP) with arrays of traps that confine one single ion or electron each \cite{paolo,irene}. For this application, they offer good arrangeability in a flat plane (2D scalability), possible miniaturization to sizes below the millimeter scale, and good accessibility with particles and radiation. Various designs and implementations of such traps for QIP with electrons exist \cite{stahl,bushev,galveexp,galvefreq,array,hex,pinder,jain,cohones,paolo,irene,gia,goldman2010penning}, probably the most straight-forward one uses concentric flat ring-shaped electrodes \cite{stahl}, which can be thought of as an axial projection of a cylindrical Penning trap onto the plane \cite{book}. This concept was proposed for QIP with 2D-arrays of planar Penning traps that hold one electron each \cite{stahl,bushev,galveexp,galvefreq,array}. Subsequent experimental work has revealed the harmonicity of the axial trap potential to be crucial for this application \cite{galveexp,bushev,irene} and has inspired more detailed considerations \cite{goldman2010penning}.

In QIP with individually confined electrons, quantum information is encoded in the spin state of the electron and its quantum state of confined motion (for details see \cite{brown1986}). For simplicity, predominantly the ground state and first excited state of the radial or axial motion are used \cite{mancini}. Processing is based on quantum-state manipulation and the entanglement with like electrons in adjacent traps \cite{lamata,ana}. This typically requires (resonant) radio-frequency or microwave excitation and detection inside the trap. The axial motion of an electron is of specific interest here since confinement conditions can be chosen such that its axial oscillation frequency and the Larmor frequency of its spin both lie in the lower microwave frequency range, presently chosen around $4\,\mathrm{GHz}$, and can hence be manipulated and detected using a microwave-only scheme. To this end, however, the axial oscillation frequency is required to be stable and independent of the electron kinetic energy (oscillation amplitude) to a high degree, for which a highly harmonic trapping potential is needed. Similar statements about trap harmonicity are true for QIP in planar radio-frequency traps \cite{hartmut}.

Planar traps are {\it by design} lacking the reflection symmetry with respect to the central plane, thus complicating their tuning for high harmonicity. There exist concepts with two identical electrode sets on co-planar surfaces facing one another to remedy this asymmetry (``mirror-image trap'' \cite{goldman2010penning,gia,fang}), however these present manufacturing issues and experimental challenges for trap loading and electron manipulation with radiation, especially on small size scales. 

It has been shown that at least four concentric electrodes with specific dimensions and gap sizes are required to allow for voltage settings that produce sufficient harmonicity \cite{goldman2010penning}. We will go into detail about the notion of ``sufficient'' harmonicity below. The four concentric electrodes are separated by three gaps, hence this design has been called a ``three-gap'' planar Penning trap in \cite{goldman2010penning}. With the use of further electrodes, the options for trap tuning increase, however so does the experimental and theoretical effort and complexity, hence we will focus on the four-electrode case which appears to be the most promising solution for real-world applications.

For trap sizes reaching the sub-mm scale, relative gap sizes generally become non-negligible \cite{auchter2023}. For the mm-scale traps considered in \cite{goldman2010penning}, the linear gap-potential approximation was found to introduce negligible corrections, while in general, finite gap and electrode sizes can produce electrostatic contributions far beyond that \cite{farrar,schmied}. 

Micro-fabricated planar Penning traps below the mm scale are already being developed and used in experiments: Sub-millimeter electrode dimensions have been demonstrated in \cite{hex}, while recent work has realized micro-fabricated surface-electrode Penning traps with electrode gaps of only a few micrometers \cite{Jain2024}. In parallel, planar Penning traps for single electrons, including the Geonium Chip, are being developed as scalable platforms for QIP and related applications \cite{Verdu2011,Pinder2017}. These developments, together with our ongoing efforts toward the experimental realization of a planar micro-Penning trap, motivate a quantitative treatment of finite electrode gaps and boundary effects in increasingly small planar Penning traps.

In the following, we first discuss the trap layout and ideal potential (Section \ref{two}), then general points to be considered in the design of a real trap (Section \ref{three}). The particular effect of gaps between electrodes and corresponding gap potentials are presented in Section \ref{gap_potentials}, and a way to quantify the resulting anharmonicities is given in Section \ref{five}. The overall strategy to optimize the potential is detailed in Section \ref{Trap_Tunability}. This is then applied under consideration of finite-boundary effects (Section \ref{boun}) and of gap effects (Section \ref{effect}).
The analytic results for optimal trap parameters are compared to finite-element simulations with COMSOL Multiphysics~\cite{comsol63} (Section~\ref{simcomp}), and then a summary of findings is given in Section \ref{summ}. We adopt the nomenclature and notation of~\cite{goldman2010penning} to ease access. 

\section{The Ideal Planar Penning Trap}
\label{two}
The planar Penning trap consists of a central circular electrode surrounded by concentric ring-shaped electrodes. These are arranged in the $(x,y)$-plane, with the axial direction $z$ perpendicular to that. We use cylindrical coordinates $(\rho,z)$ with the radial coordinate $\rho$ given by $\rho^2=x^2+y^2$ and the axial coordinate $z$. The electrode radii $\rho_i$ are measured from the center at $\rho=(x,y)=0$ to the mid of the gap between electrode $i$ and $i+1$, except for the outermost electrode where it is measured to the outer edge. Each electrode $i$ is set to a specific constant voltage $V_i$, with the outermost electrode at ground potential. In addition to the electrostatic potential, the trap uses a homogeneous magnetic field applied in $z$ - direction for electron confinement in the radial direction, in full similarity to traditional Penning traps, and we will not be concerned with this, since the axial potential (and thus the axial motion) is independent of this field.

\begin{figure}[ht]
    \centering
    \includegraphics[width=\columnwidth]{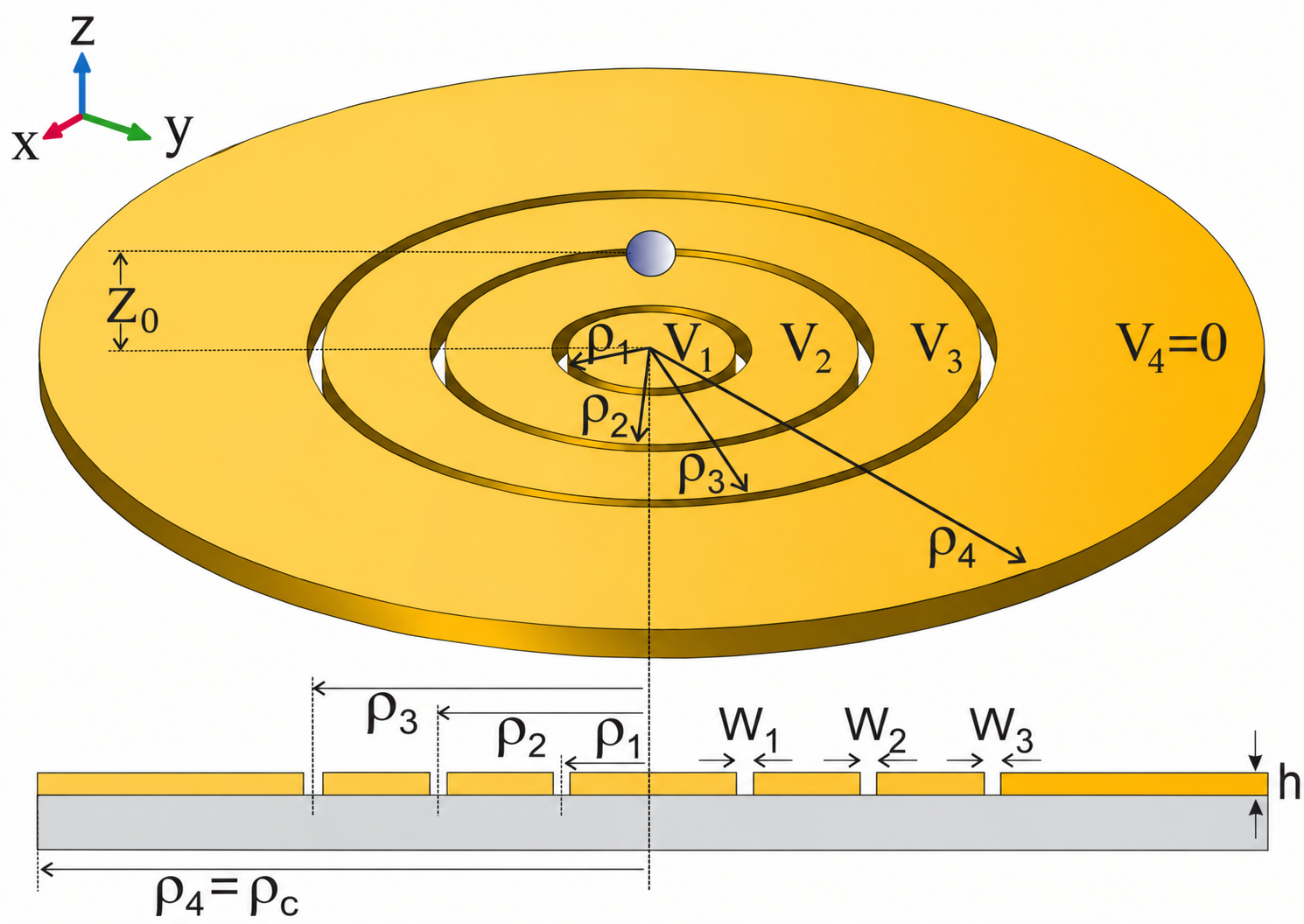}
    \caption{Top: Electrode structure of a four-electrode (three-gap) planar Penning trap. A confined electron is indicated in the trap center at a height $z_0$ above the plane of the electrodes. Bottom: Cross-sectional view. The electrodes are supported by a carrier substrate and are separated by gaps of widths $w_i$ and height $h$ for electrical insulation from one another.\label{trap} }
\end{figure}

Between electrodes, gaps of width $w_i$ are required for electric insulation from each other. Ideally, these gaps would have zero width, which of course is infeasible in reality, and the consequences of this will be discussed in the following. The resulting electrostatic potential $V(\rho,z)$ in the space above the plane ($z>0$) is the linear superposition (sum) of the contributions from each electrode and gap. A schematic of the arrangement for four electrodes (three gaps) is shown in Fig.~\ref{trap}.  

We wish the resulting electrostatic potential $V(\rho,z)$ in the space above the plane to be the quadrupole potential of the ideal Penning trap given by $V(\rho, z) \propto (z - z_0)^2 - \rho^2/2$. Since we are interested only in the axial motion, the problem is reduced to the potential on the central axis by setting $\rho=0$, so that the ideal axial potential has the shape $V(z) \propto (z-z_0)^2$.

When we look at an arbitrary axial trap potential and write down a Taylor expansion around $z=z_0$, this results in $V(z)=V(z_0) + V'(z_0)(z-z_0) + V''(z_0)(z-z_0)^2/2+\dots$, where the prime sign indicates a spatial derivative with respect to $z$. In this sum, the first two terms (offset and slope) are irrelevant for the frequency of the axial motion \cite{brown1986}, while the third term measures the spatial curvature $V_0/\rho_1^2$ of the potential at $z=z_0$ and thus defines the axial oscillation frequency $\omega_z$ of the electron. Higher-order terms (fourth term and beyond) introduce contributions to the axial frequency that are position-dependent, and thus undesired. When those are zero, we find the ideal axial trap potential
\begin{equation}
V(z) = \frac{1}{2} \frac{V_0}{\rho_1^2} (z - z_0)^2, \label{eq1}
\end{equation}
which leads to an axial oscillation frequency $\omega_z$ of a trapped electron given by
\begin{equation}
\omega_z = \sqrt{\frac{e}{m} \frac{V_0}{\rho_1^2}}, \label{eq:axial_freq1}
\end{equation}
where $e$ and $m$ are the charge and mass of the electron, respectively. The potential curvature $V_0/\rho_1^2$ follows from the choice of all geometry and voltage parameters $(\rho_i,w_i,V_i)$, as will be discussed below. Particularly, this expression for $\omega_z$ shows that in the ideal case, the axial frequency remains independent of the axial oscillation amplitude (energy) of the particle, which is the desired feature for the present applications.

\section{Design considerations towards a real planar Penning trap}
\label{three}
\subsection{Tunability of the Electrostatic Potential}
Considering effort, complexity and reliability of operation, we desire to work with the smallest number of electrodes possible. The simplest case, a two-electrode (single-gap) trap, cannot produce a potential minimum for any choice of size and voltage parameters and can thus not be a trap \cite{stahl,goldman2010penning}. A three-electrode (two-gap) trap can feature such a minimum, but the harmonicity of the potential around the minimum cannot be sufficiently tuned and hence it is not useful for the applications under discussion \cite{goldman2010penning}. By use of four electrodes (three gaps), tuning to a sufficient degree is possible, actually in more than one way, and hence this is the case we will focus on. Beyond four electrodes, these possibilities increase further, but at the cost of considerable additional complexity.

\subsection{Electrostatic Boundary}
The ideal case as discussed above needs to be complemented by a number of considerations when a realistic trap is to be designed. First, the determination of the trap's electrostatic properties requires boundary conditions in addition to the definitions shown in Fig.~\ref{trap}. 
Commonly, the easiest choice is a boundary with ground potential $V \equiv 0$ at infinity from the trap center. In reality, it is of course impossible to place the ground of the experimental setup infinitely far away from the trap electrodes. As a result, finite-boundary effects -- both radial and axial -- become relevant. A convenient and widely used approach to account for these boundary conditions is to enclose the trap within a grounded conducting cylinder of inner radius $\rho_c$ and height $z_c$, closed by a flat plate (see also Fig.~12 of Ref.~\cite{goldman2010penning}).

\subsection{Working Point of the Trap}
\label{working_point}
\begin{figure*}[t]
    \centering
    \includegraphics[width=0.85\columnwidth]{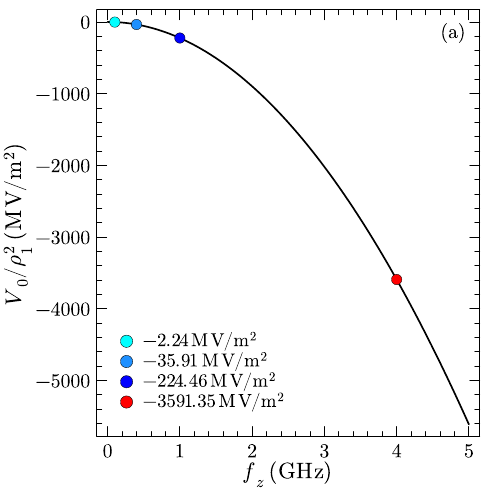}
    \includegraphics[width=0.85\columnwidth]{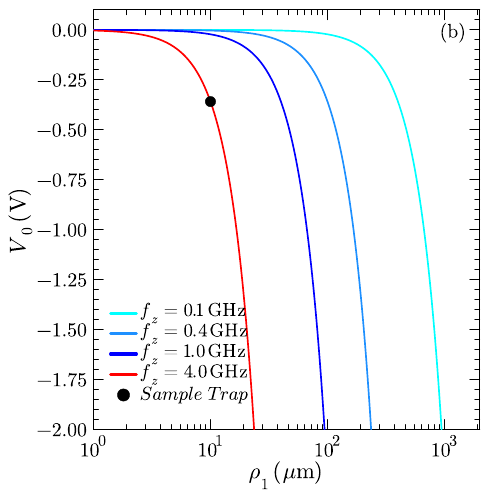}
    \caption{
    (a) Required electrostatic curvature $V_{0}/\rho_{1}^{2}$ as a function of the desired axial frequency $f_z$, calculated using Eq.~\eqref{eq:axial_freq1}. The marked points correspond to axial frequencies of 0.1, 0.4, 1.0, and 4.0~GHz, respectively.
    (b) Required trap depth $V_{0}$ as a function of the trap size $\rho_{1}$ for the corresponding axial frequencies. The solid circle indicates the selected parameters, $V_0=-0.3591\,\mathrm{V}$ and $\rho_1=10.0\,\mu\mathrm{m}$, corresponding to $f_z=4.0\,\mathrm{GHz}$.
    }
    \label{fig:curvature_fz}
\end{figure*}
For a given application, the design choices may either be driven by size requirements or by the desire to work at a specific axial frequency $\omega_z$. In principle, any value of the axial frequency can be realised at any size scale $\rho_1$ of the trap, however in reality one is limited by the magnitude of the trap depth $V_0$ that can reliably be created from the applied electrode voltages $V_i$. Practical voltage limits are imposed by the finite resistances between electrodes and the availability of sufficiently precise and stable voltage supplies, currently the typical limits are of the order of tens of Volts. Similarly, the characteristic radius $\rho_{1}$ is constrained by practical fabrication limits and minimum feature sizes which will be discussed below. For the QIP schemes mentioned above \cite{mancini,lamata,ana}, this frequency is typically chosen such that at the given temperature $T$ of the axial electron  motion, the probability $P=(1-e^{-\kappa})e^{-N_z \kappa}$ with $\kappa=\hbar \omega_z/(k_BT)$ of being in the $N_z=0$ ground state of the level scheme \cite{brown1986} is sufficiently close to unity.

Eq.~\eqref{eq:axial_freq1} gives the axial frequency $\omega_z$ as a function of the electrostatic potential curvature $V_0/\rho_1^2$. Fig.~\ref{fig:curvature_fz}(a) shows the potential curvature $V_0/\rho_1^2$ as a function of the desired axial oscillation frequency $f_z=\omega_z/(2\pi)$, calculated for an electron. Four specific values of the potential curvature, corresponding to $f_z=0.1\,\mathrm{GHz}$, $0.4\,\mathrm{GHz}$, $1.0\,\mathrm{GHz}$, and $4.0\,\mathrm{GHz}$, are indicated as examples. Fig.~\ref{fig:curvature_fz}(b) shows the required trap depth $V_0$ as a function of the trap size $\rho_1$ for these frequencies. The solid circle marks the selected design parameters for the sample trap, $V_0=-0.3591\,\mathrm{V}$ and $\rho_1=10.0\,\mu\mathrm{m}$, corresponding to an axial frequency of $f_z=4.0\,\mathrm{GHz}$.

\subsection{Required Optimisation of the Trap Potential}
\label{reqopt}
The use of a single electron in a planar Penning trap for QIP requires the axial oscillation frequency to be sufficiently constant to spectrally resolve a single electron and its axial frequency shifts associated with quantum-state transitions in information processing \cite{peil1999quantum}. In particular, even for perfectly stable electrode voltages, the finite temperature associated with the axial electron motion will lead to a shift and broadening of the detected frequency spectrum if the axial potential is anharmonic \cite{brown1986}. 
The intrinsic spectral width of resonant axial detection by a tuned electronic circuit is given by 
\begin{equation}
    \gamma_z =
    \frac{e^2 R}{m D_1^{\,2}},
    \label{eq:gamma_z}
\end{equation}
where $R$ is the on-resonance resistance of the detection circuit, and the quantity $D_1$ represents the effective distance of the electrode used for signal pickup as seen by the axially oscillating electron \cite{book}. The quality factor of such a detector is $Q=f_z/\gamma_z$ and is most often in the range from roughly $10^3$ to $10^5$, given by technical details \cite{ulm}.

In a real planar trap with its residual anharmonicities, the axial frequency $\omega_z$ depends on the axial oscillation amplitude $A$ as will be discussed in detail in Sec.~\ref{sec:anharmonic_frequency_link}. Hence, a thermal distribution of axial energies leads to a corresponding distribution of axial oscillation amplitudes and thus to a finite axial frequency width $\sigma_f$, see Eq.~\eqref{eq:thermal_spread}. It will be the main goal of our optimization to minimize $\sigma_f$ by appropriate choice of geometry and voltage parameters of the trap.
If the trap is not sufficiently optimized, the thermally induced broadening $\sigma_f$ can exceed the detection width $\gamma_z$, preventing single-electron detection. The goal of optimization is thus $\sigma_f \ll \gamma_z$.

\subsection{Fabrication Limits on Electrode Gaps}

For micro-fabricated planar trap designs, the achievable electrode geometry is constrained by the capabilities of the fabrication process and by electrical breakdown at small electrode gaps. Surface-electrode architectures based on planar electrodes have been demonstrated using standard micro-fabrication techniques \cite{chiaverini2005surface,seidelin2006microfabricated}. Micrometer-scale electrode gaps have been demonstrated in fabricated surface-electrode ion traps. In particular, a CMOS- (complementary metal–oxide–semiconductor)-fabricated ion trap employed electrode gaps of approximately $5\,\mu\mathrm{m}$ \cite{mehta2014ion}.
Advances in semiconductor and micro-electro-mechanical-systems (MEMS) fabrication have enabled increasingly small and precisely defined electrode features \cite{jaeger2002introduction,madou2011fundamentals,romaszko2020engineering}. In particular, ion traps have been fabricated using commercial $90\,\mathrm{nm}$ CMOS foundry processes, demonstrating the compatibility of surface-electrode ion traps with advanced semiconductor fabrication \cite{mehta2014ion}. These developments demonstrate that micrometer-scale electrode gaps are experimentally achievable, while sub-micron features can also be realized using contemporary semiconductor fabrication processes.

\section{Axial trap Potential in the presence of electrode gaps}
\label{gap_potentials}

For this discussion, we introduce scaled variables that make the formulation dimensionless. We scale all potentials by the trap depth $V_0$, i.e. we have scaled applied voltages $\tilde{V}_i = V_i/V_0$ and a scaled trap potential $\tilde{V} = V/V_0$. Likewise, we scale the radial and axial dimensions by the radius $\rho_1$, i.e. we have $\tilde{\rho} = \rho/\rho_1$ and $\tilde{z} = z/\rho_1$. Throughout this work, the gap widths between all electrodes are considered identical. Then the geometry of an $N$-gap planar trap is given by a set $\{\tilde{\rho_i}\} = \{1, \tilde{\rho_2}, \tilde{\rho_3},....,\tilde{\rho_N}, \tilde{\rho}_{N+1} = \tilde{\rho_c}\}$.

Considering the rotational symmetry about the $z$-axis, the potential of a planar Penning trap for infinitesimal gap widths is a linear superposition of the contributions $\phi_i$ from the individual electrodes only, and depends linearly on the relative applied voltages $V_i$. It is hence given by the sum
\begin{equation}
    {V}(\tilde{\rho}, \tilde{z}) = \sum_{i=1}^{N} V_i\phi_i(\tilde{\rho}, \tilde{z}), \label{eq:3D_potential}
\end{equation}
where the functions $\phi_i$ are solutions of Laplace’s equation subject to specific boundary conditions and are independent of the applied electrode voltages, depending only on the geometry parameters of the trap. For non-negligible gap widths, however, this is not sufficient.

To introduce the effect of gaps, it is necessary to know the exact potential variation within each gap at $\tilde{\rho} = \tilde{\rho_i}$. For this, we define a gap potential of the $i^{th}$ gap in the $z = 0$ plane as $V_{g}^i(\tilde{\rho},0)$. We will consider different types of gap potentials, namely
\begin{itemize}
    \item infinitesimal gaps, i.e. gaps with assumed zero width but insulating property (A)
    \item linear gap potential (B)
    \item gap potential according to the corner-law (C)
    \item cubic gap potential (D), 
\end{itemize}
to evaluate their overall effect on the potential in the vicinity of the equilibrium position $\tilde{z}_0$. 

Under azimuthal symmetry, Laplace’s equation in cylindrical coordinates \cite{Jackson1999} admits separable solutions with radial and axial components. Physical constraints requiring the potential to remain finite at the origin exclude divergent Bessel-function solutions, while the planar trap radial boundary condition enforces oscillatory radial behavior. Consequently, the electrostatic potential can be expressed as a Fourier–Bessel expansion \cite{Jackson1999, Kusse2006}. The full derivation, following standard treatments, is also given in Appendix~A of \cite{Goldman2011}, yielding
\begin{equation}
    V(\tilde{\rho},\tilde{z}) 
    = \sum_{k} \left( A_k e^{-k\tilde{z}} 
    +
    B_k e^{k\tilde{z}} \right) J_0(k\tilde{\rho}),
    \label{eq:Sol_Bessels}
\end{equation}
where the coefficients $A_k$ and $B_k$ are to be determined using the appropriate boundary conditions. A planar Penning trap with the boundaries at finite distances has the following boundary conditions:
\begin{eqnarray}
    V(\tilde{\rho}=\tilde{\rho_c}, \tilde{z}) = 0, \label{eq:rad_bound} \\
    V(\tilde{\rho}, \tilde{z} = \tilde{z_c}) = 0, \label{eq:axial_bound} \\
    V(\tilde{\rho}, 0) = V(\tilde{\rho}).   \label{eq:electrode_plane_bound}
\end{eqnarray}
The radial boundary condition in Eq.~\eqref{eq:rad_bound} restricts the separation constant to discrete values determined by
\begin{equation}
    J_0(k\tilde{\rho}_c)=0 \;\;\Rightarrow\;\; k=k_n=\frac{\alpha_{0n}}{\tilde{\rho}_c},
    \label{eq:bessel_cond}
\end{equation}
where \(\alpha_{0n}\) denotes the \(n^{\text{th}}\) zero of \(J_0(x)\). The resulting potential can therefore be written as
\begin{equation}
    V(\tilde{\rho},\tilde{z})
    =
    \sum_{n=1}^{\infty}\left(A_n e^{-k_n \tilde{z}}
    +
    B_n e^{k_n \tilde{z}}\right)J_0(k_n \tilde{\rho}).
    \label{eq:Sol_Coeff_Bessels}
\end{equation}
The remaining two boundary conditions in Eqs.~\eqref{eq:axial_bound} and~\eqref{eq:electrode_plane_bound}, combined with Bessel-function orthogonality, uniquely determine both coefficients. The resulting coefficients are
\begin{equation}
    A_n =
    \frac{e^{k_n \tilde{z}_c}}{{\tilde{\rho}_{c}}^2 J_1^2(\alpha_{0n}) \sinh(k_n \tilde{z}_c)}
    \int_{0}^{\tilde{\rho}_c}
    \tilde{\rho}\,
    J_0(k_n \tilde{\rho})\,
    V(\tilde{\rho})\,
    \mathrm{d}\tilde{\rho}
    \label{eq:Coeff_An}
\end{equation}
and
\begin{equation}
    B_n =
    \frac{-e^{-k_n \tilde{z}_c}}{{\tilde{\rho}_{c}}^2 J_1^2(\alpha_{0n}) \sinh(k_n \tilde{z}_c)}
    \int_{0}^{\tilde{\rho}_c}
    \tilde{\rho}\,
    J_0(k_n \tilde{\rho})\,
    V(\tilde{\rho})\,
    \mathrm{d}\tilde{\rho}.
    \label{eq:Coeff_Bn}
\end{equation}
Now inserting Eqs.~\eqref{eq:Coeff_An} and~\eqref{eq:Coeff_Bn} into Eq.~\eqref{eq:Sol_Coeff_Bessels} yields the trap potential for $\tilde{z} \geq 0$
\begin{multline}
    V(\tilde{\rho}, \tilde{z})
    =
    \sum_{n=1}^{\infty}
    \frac{-2}{\tilde{\rho}_{c}^{2}J_{1}^{2}(\alpha_{0n})}
    \frac{\sinh\!\big(k_n(\tilde{z}-\tilde{z}_c)\big)}
         {\sinh(k_n\tilde{z}_c)}
    \\
    \qquad\times
    \left[
    \int_{0}^{\tilde{\rho}_c}
    \tilde{\rho}\,
    J_0(k_n\tilde{\rho})\,
    V(\tilde{\rho})\,
    \mathrm{d}\tilde{\rho}
    \right]
    J_0(k_n\tilde{\rho}).
    \label{eq:Pot_General_Sol}
\end{multline}
Consequently, once the potential distribution $V(\tilde{\rho})$ on the $z=0$ plane is specified, the corresponding trap potential for $\tilde{z} \geq 0$ follows directly. In the following subsections, different choices of $V(\tilde{\rho})$ -- corresponding to infinitesimal and finite gap widths with various gap-potential models -- are considered and substituted into Eq.~\eqref{eq:Pot_General_Sol}.

\subsection{Infinitesimal Gaps}
\label{Infinitesimal_Gaps}
We first consider the infinitesimal-gap limit, in which the gaps between adjacent electrodes are assumed to have zero width (but of course still insulate the electrode voltages from one another). In this case, the electrostatic potential in the \(z = 0\) plane is piecewise constant. When the \(i^{\text{th}}\) electrode is held at unit potential and all other electrodes are grounded, the corresponding boundary potential is given by
\begin{equation}
    \phi_i(\tilde{\rho}, 0) =
    \begin{cases}
    0, & \text{for } \tilde{\rho} < \tilde{\rho}_{i-1}, \\
    1, & \text{for } \tilde{\rho}_{i-1} \le \tilde{\rho} \le \tilde{\rho}_i, \\
    0, & \text{for } \tilde{\rho} > \tilde{\rho}_i .
    \end{cases}
    \label{eq:electode_plane_bound_zero_gap}
\end{equation}
The total potential applied in the \(z = 0\) plane is then obtained by linear superposition of the individual electrode contributions, such that we have
\begin{equation}
    V(\tilde{\rho}) = \sum_{i=1}^{N} V_i \, \phi_i(\tilde{\rho}, 0).
    \label{eq:radial_pot_zero_gap}
\end{equation}
Substituting the boundary potential of Eq.~\eqref{eq:radial_pot_zero_gap}, with the individual electrode contributions defined by Eq.~\eqref{eq:electode_plane_bound_zero_gap}, into the general solution of Eq.~\eqref{eq:Pot_General_Sol}, and evaluating the resulting integral gives
\begin{multline}
    V(\tilde{\rho}, \tilde{z})
    =
    \sum_{i=1}^{N}\Delta V_i
    \sum_{n=1}^{\infty}
    \frac{2}{\tilde{\rho}_{c}^{2}J_{1}^{2}(\alpha_{0n})}
    \frac{\sinh\!\big(k_n(\tilde{z}-\tilde{z}_c)\big)}
         {\sinh(k_n\tilde{z}_c)}
    \\
    \qquad\times
    \left[
    \frac{\tilde{\rho}_i}{k_n}
    J_1(k_n\tilde{\rho}_i)
    \right]
    J_0(k_n\tilde{\rho}).
    \label{eq15}
\end{multline}
which is equivalent to Eq.~(55) of Ref.~\cite{goldman2010penning} upon substituting Eq.~(56) of the same reference and using the conventions $\tilde{\rho}_0=0$, $V_{N+1}=0$, and $\Delta V_i=V_{i+1}-V_i$.

On the symmetry axis $\tilde{\rho} = 0$, where $J_0(k_n\tilde{\rho}) = 1$, the axial potential of an $N$-gap trap is then given by 
\begin{equation}
    V(0, \tilde{z}) =
    \sum_{i=1}^{N}\Delta{V_i}\Phi_i^{(0)}(\tilde{z}; \tilde{\rho_c}, \tilde{z_c}),
    \label{eq:Vz_gapless}
\end{equation}
where dimensionless potentials are given by 
\begin{multline}
    \Phi_i^{(0)}(\tilde{z};\tilde{\rho}_c,\tilde{z}_c)
    =
    \sum_{n=1}^{\infty}
    \frac{2}
    {\tilde{\rho}_{c}^{2}J_{1}^{2}(\alpha_{0n})}
    \frac{\sinh\!\big(k_n(\tilde{z}-\tilde{z}_c)\big)}
         {\sinh(k_n\tilde{z}_c)}
    \\
    \times
    \left[
    \frac{\tilde{\rho}_i}{k_n}
    J_1(k_n\tilde{\rho}_i)
    \right].
    \label{eq:Laplace_Sol_Gapless}
\end{multline}
It can be seen that these potentials are independent of the voltages applied to the trap and depend solely on the relative geometric parameters of the trap electrodes. Consequently, most of the essential properties of a planar Penning trap can be inferred directly from the on-axis potential. 

In the infinite external boundary limit ($\tilde{\rho}_c, \tilde{z}_c \rightarrow \infty$), the Laplace solution of Eq.~\eqref{eq:Laplace_Sol_Gapless} simplifies considerably. In this limit, the discrete eigenvalues $k_n = \alpha_{0n}/\tilde{\rho}_c$ form a continuous spectrum as $\tilde{\rho}_c \to \infty$, and the summation over modes transforms into an integral representation. As $\tilde{z}_c \to \infty$, the axial boundary recedes, which simplifies the hyperbolic dependence to an exponentially decaying form. As a result, the bounded-domain solution reduces to the corresponding free-space Laplace solution,
\begin{equation}
    \Phi_i^{(0)}(\tilde{z})
    =
    \frac{\tilde{z}}{\sqrt{\tilde{\rho}_i^2 + \tilde{z}^2}} - 1.
    \label{eq:Laplace_Sol_Gapless_infinite_bound}
\end{equation}

\subsection{Linear Gap Model}
\label{Linear_Model}
\begin{figure}[t]
    \centering
    \includegraphics[width=\columnwidth]{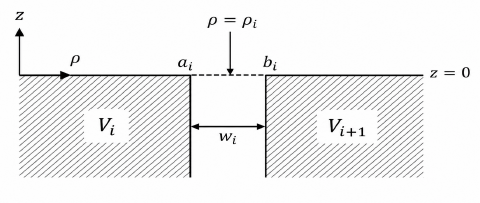}
    \caption{Schematic of the \(i^{\text{th}}\) electrode gap of width \(w_i\), centered at radius \(\rho = \rho_i\).}
    \label{ith_gap}
\end{figure}

To account for finite gap widths, we first adopt a linear gap-potential model. In this approach, the contribution of the gap to the overall trap potential is approximated by a uniform gap field, which provides an adequate description when the widths and heights of the adjacent electrodes are large compared to the gap width. This approximation is equivalent to modeling the gap field as that of a semi-infinite gap while neglecting edge effects, as illustrated in Fig.~\ref{ith_gap}.
The electrodes that neighbour the $i^{th}$ gap of width $\tilde{w}_i$ are biased with voltages $\tilde{V_i}$ and $\tilde{V}_{i+1}$, respectively. Under this linear approximation, the gap potential of the $i^{th}$ gap at radius $\tilde{\rho} = \tilde{\rho_i} $ in the $z=0$ plane is given by 
\begin{equation}
    {V_g^i}(\frac{\tilde{\xi}_i}{\tilde{w}_i}) = {A}_0 + {A}_1 (\frac{\tilde{\xi}_i}{\tilde{w}_i}), 
    \label{eq:Lin_gap_pot}
\end{equation}
where we define a shifted variable $\tilde{\xi}_i = \tilde{\rho} - \tilde{a}_i$ and $\tilde{a}_i = \tilde{\rho}_i - \tilde{w}_i/2$. The constants ${A}_0$ and ${A}_1$ are determined by the following electrode plane boundary conditions
\begin{equation}
    {V_g^i}(\frac{\tilde{\xi}_i}{\tilde{w}_i}) =
    \begin{cases}
    V_i, & \text{for } \frac{\tilde{\xi}_i}{\tilde{w}_i} = 0, \\
    V_{i+1}, & \text{for } \frac{\tilde{\xi}_i}{\tilde{w}_i} = 1, \\
    \frac{V_{i} + V_{i+1}}{2}, & \text{for } \frac{\tilde{\xi}_i}{\tilde{w}_i} = 1/2.  
    \label{eq19}
    \end{cases}
\end{equation}
For the linear gap model, the two endpoint potential conditions determine the coefficients \(A_0=V_i\) and \(A_1=\Delta V_i\), where \(\Delta V_i=V_{i+1}-V_i\). The midpoint-potential condition is automatically satisfied by linear interpolation between the endpoint voltages. Moreover, the electric field is uniform throughout the gap, so the endpoint field-symmetry condition is also automatically satisfied:
\begin{equation}
\vec{E}(0)=\vec{E}(1).
\label{eq20}
\end{equation}
Thus, neither the midpoint-potential condition nor the endpoint field-symmetry condition provides an additional independent constraint for this model.

Substituting the coefficients into Eq.~\eqref{eq:Lin_gap_pot} gives the linear gap potential
\begin{equation}
    {V_g^i}(\tilde{\rho}, 0) = V_i + \Delta{V_i}(\frac{\tilde{\rho} - \tilde{a}_i}{\tilde{w}_i}). \label{eq21}
\end{equation}
The total potential in the $z=0$ plane including the linear gap potential is given by
\begin{equation}
    {V}(\tilde{\rho}) 
    = 
    \sum_{i=1}^{N} V_i \phi_i(\tilde{\rho}, 0) 
    + 
    \sum_{i=1}^{N}[V_i\phi_i^{g(1)}(\tilde{\rho}, 0) + \Delta{V_i}\phi_i^{g(2)}(\tilde{\rho}, 0)], \label{eq:Lin_rad_gap_pot}
    \end{equation}
    where the electrode and gap boundary conditions to calculate each $\phi_i$ in Eq.~\eqref{eq:Lin_rad_gap_pot} are as follows and also inferred from Fig.~\ref{ith_gap};
    \begin{equation}
    \phi_i(\tilde{\rho}, 0) =
    \begin{cases}
    0, & \text{for } \tilde{\rho} < \tilde{b}_{i-1}, \\
    1, & \text{for } \tilde{b}_{i-1} \le \tilde{\rho} \le \tilde{a}_i, \\
    0, & \text{for } \tilde{\rho} > \tilde{a}_i .
    \end{cases}
    \label{eq:electrode_bound_with_lin_gap}
\end{equation}
\begin{equation}
    \phi_i^{g(1)}(\tilde{\rho}, 0) =
    \begin{cases}
    0, & \text{for } \tilde{\rho} < \tilde{a}_{i}, \\
    1, & \text{for } \tilde{a}_{i} \le \tilde{\rho} \le \tilde{b}_i, \\
    0, & \text{for } \tilde{\rho} > \tilde{b}_i.
    \label{eq:gap_bound_with_lin_gap1}
    \end{cases}
\end{equation}
\begin{equation}
    \phi_i^{g(2)}(\tilde{\rho}, 0) =
    \begin{cases}
    0, & \text{for } \tilde{\rho} < \tilde{a}_{i}, \\
    \frac{\tilde{\rho}-\tilde{a}_i}{\tilde{w}_i}, & \text{for } \tilde{a}_{i} \le \tilde{\rho} \le \tilde{b}_i, \\
    0, & \text{for } \tilde{\rho} > \tilde{b}_i.
    \label{eq:gap_bound_with_lin_gap2}
    \end{cases}
\end{equation}
Substituting the boundary potential from Eq.~\eqref{eq:Lin_rad_gap_pot}, with the electrode and gap contributions defined by Eqs.~\eqref{eq:electrode_bound_with_lin_gap}--\eqref{eq:gap_bound_with_lin_gap2}, into the general solution of Eq.~\eqref{eq:Pot_General_Sol}, and evaluating the resulting integrals using the conventions $\tilde{b}_0=0$, $V_{N+1}=0$, and $\tilde{b}_i=\tilde{\rho}_i+\tilde{w}_i/2$, yields the trap potential for the linear gap model as
\begin{multline}
    V(\tilde{\rho},\tilde{z})
    =
    \sum_{i=1}^{N}\frac{\Delta V_i}{\tilde{w}_i}
    \sum_{n=1}^{\infty}
    \frac{2}
    {\tilde{\rho}_{c}^{2}J_{1}^{2}(\alpha_{0n})}
    \frac{\sinh\!\big(k_n(\tilde{z}-\tilde{z}_c)\big)}
         {\sinh(k_n\tilde{z}_c)}
    \\
    \times
    \left[
    \frac{\tilde{a}_i}{k_n^{2}}J_0(k_n\tilde{a}_i)
    -
    \frac{\tilde{b}_i}{k_n^{2}}J_0(k_n\tilde{b}_i)
    \right.
    \\
    \left.
    +
    \frac{1}{k_n^{2}}
    \int_{\tilde{a}_i}^{\tilde{b}_i}
    J_0(k_n\tilde{\rho})\,
    \mathrm{d}\tilde{\rho}
    \right]
    J_0(k_n\tilde{\rho}).
    \label{eq:Trap_3D_pot_Lin}
\end{multline}
On the symmetry axis \(\tilde{\rho}=0\), where \(J_0(k_n\tilde{\rho})=1\), the axial potential of an \(N\)-gap trap reduces to
\begin{equation}
    V(0, \tilde{z}) =
    \sum_{i=1}^{N}\Delta {V_i}\,
    \Phi_i^{lin}(\tilde{z}; \tilde{w}_i, \tilde{\rho}_c, \tilde{z}_c),
    \label{eq:Vz_Lin}
\end{equation}
where the corresponding dimensionless potential functions are defined as
\begin{multline}
    \Phi_i^{\mathrm{lin}}
    (\tilde{z};\tilde{w}_i,\tilde{\rho}_c,\tilde{z}_c)
    =
    \frac{1}{\tilde{w}_i}
    \sum_{n=1}^{\infty}
    \frac{2}
    {\tilde{\rho}_{c}^{2}J_{1}^{2}(\alpha_{0n})}
    \frac{\sinh\!\big(k_n(\tilde{z}-\tilde{z}_c)\big)}
         {\sinh(k_n\tilde{z}_c)}
    \\
    \times     
    \left[
    \frac{\tilde{a}_i}{k_n^{2}}
    J_0(k_n\tilde{a}_i)
    -
    \frac{\tilde{b}_i}{k_n^{2}}
    J_0(k_n\tilde{b}_i)
    \right.
    \\
    \left.
    +
    \frac{1}{k_n^{2}}
    \int_{\tilde{a}_i}^{\tilde{b}_i}
    J_0(k_n\tilde{\rho})\,
    \mathrm{d}\tilde{\rho}
    \right].
    \label{eq:Laplace_Lin}
\end{multline}

\subsection{Corner-Effect Gap Model}
\label{Corner-Law_Model}
Further improvement over the linear gap-potential model can be achieved by incorporating higher-order terms in the gap potential. In Ref.~\cite{farrar}, a fifth-order polynomial approximation was employed to describe the potential variation across the gap. A more physically realistic description is obtained by explicitly accounting for the electric-field singularity at the electrode corners. To this end, the semi-infinite gap geometry shown in Fig.~\ref{ith_gap} is again considered, with the widths and heights of the adjacent electrodes assumed to be much larger than the gap width. Under these conditions, and following the analysis of the electric field near a conducting corner~\cite{Jackson1999,farrar}, the gap potential along the radial direction in the plane defined by $z=0$ is given by
\begin{multline}
    V_g^i\!\left(\frac{\tilde{\xi}_i}{\tilde{w}_i}\right)
    =
    V_k
    +
    \sum_{n=1}^{\infty}
    \Biggl[
    c_n
    \left(
    \frac{\tilde{\xi}_i}{\tilde{w}_i}
    \right)^{2n/3}
    \\
    +
    d_n
    \left(
    1-
    \frac{\tilde{\xi}_i}{\tilde{w}_i}
    \right)^{2n/3}
    \Biggr],
    \label{eq29}
\end{multline}
where the coefficients are given by
\[
    c_n = a_n\, \tilde{w}_i^{2n/3}\, \sin\!\left(\frac{n\pi}{3}\right), \qquad
    d_n = b_n\, \tilde{w}_i^{2n/3}\, \sin\!\left(\frac{n\pi}{3}\right).
\]
The offset voltage \( V_k \) remains undetermined and depends on the potential far from the edge of the electrode. Consequently, near the edge of the electrode \( \tilde{\xi}_i = 0 \), only the first term in the series is significant. Therefore, the gap potential of the \( i^{\mathrm{th}} \) gap can be approximated as
\begin{equation}
    V_g^i\!\left(\frac{\tilde{\xi}_i}{\tilde{w}_i}\right)
    =
    V_k
    +
    \left[
    {c_1} \left( \frac{\tilde{\xi}_i}{\tilde{w}_i}\right)^{2/3} 
    \!\!\!\!\!\!+
    {d_1} \left( 1 - \frac{\tilde{\xi}_i}{\tilde{w}_i}\right)^{2/3}
    \right].
    \label{eq30}
\end{equation}
Applying the three potential conditions in Eq.~\eqref{eq19} determines the coefficients as \(V_k=(V_i+V_{i+1})/2\) and \(d_1=-c_1=-\Delta V_i/2\), where \(\Delta V_i=V_{i+1}-V_i\). Substituting these coefficients into Eq.~\eqref{eq30} yields the first-order corner-effect gap potential:
\begin{multline}
    V_g^i\!\left(\frac{\tilde{\xi}_i}{\tilde{w}_i}\right)
    =
    V_i
    +\frac{\Delta V_i}{2}
    \\
    \times
    \left[
    1
    +
    \left(
    \frac{\tilde{\xi}_i}{\tilde{w}_i}
    \right)^{2/3}
    -
    \left(
    1-
    \frac{\tilde{\xi}_i}{\tilde{w}_i}
    \right)^{2/3}
    \right].
    \label{eq31}
\end{multline}
Substituting the corner-effect gap potential of Eq.~\eqref{eq31} into Eq.~\eqref{eq:Lin_rad_gap_pot}, and applying the boundary conditions specified in Eqs.~\eqref{eq:electrode_bound_with_lin_gap}--\eqref{eq:gap_bound_with_lin_gap1}, together with the corresponding definition of $\phi_i^{g(2)}(\tilde{\rho},0)$ given by
\begin{equation}
    \phi_i^{g(2)}(\tilde{\rho}, 0)
    =
    \begin{cases}
    0, & \tilde{\rho} < \tilde{a}_{i}, \\[4pt]
    \displaystyle
    \phi_{ab}^{c},
    & \tilde{a}_{i} \le \tilde{\rho} \le \tilde{b}_i, \\[8pt]
    0, & \tilde{\rho} > \tilde{b}_{i},
    \end{cases}
    \label{eq32}
\end{equation}
with
\begin{equation} \nonumber
   \phi_{ab}^{c}
   = 
  \frac{1}{2}
    \left[
        1
        + \left(\frac{\tilde{\rho}-\tilde{a}_i}{\tilde{w}_i}\right)^{2/3}
        - \left(1-\frac{\tilde{\rho}-\tilde{a}_i}{\tilde{w}_i}\right)^{2/3}
    \right],
\end{equation}
into the general solution of Eq.~\eqref{eq:Pot_General_Sol} and evaluating the resulting integral yields the axial potential
\begin{equation}
    V(0, \tilde{z}) =
    \sum_{i=1}^{N}
    \Delta {V}_i\,
    \Phi_i^{c}(\tilde{z}; \tilde{w}_i, \tilde{\rho}_c, \tilde{z}_c),
    \label{eq:Vz_corner}
\end{equation}
where dimensionless potentials are given by
\begin{multline}
    \Phi_i^c(\tilde{z};\tilde{w}_i,\tilde{\rho}_c,\tilde{z}_c)
    =
    \sum_{n=1}^{\infty}
    \frac{2}
    {\tilde{\rho}_{c}^{2}J_{1}^{2}(\alpha_{0n})}
    \frac{\sinh\!\big(k_n(\tilde{z}-\tilde{z}_c)\big)}
         {\sinh(k_n\tilde{z}_c)}
    \\
    \times
    \Bigg[
    \frac{\tilde{a}_i}{2k_n}
    J_1(k_n\tilde{a}_i)
    +
    \frac{\tilde{b}_i}{2k_n}
    J_1(k_n\tilde{b}_i)
    \\
    -
    \frac{1}{2}
    \int_{\tilde{a}_i}^{\tilde{b}_i}
    \Bigg\{
    \left(
    \frac{\tilde{\rho}-\tilde{a}_i}{\tilde{w}_i}
    \right)^{2/3}
    -
    \left(
    1-
    \frac{\tilde{\rho}-\tilde{a}_i}{\tilde{w}_i}
    \right)^{2/3}
    \Bigg\}
    \\
    \times
    \tilde{\rho}\,
    J_0(k_n\tilde{\rho})\,
    \mathrm{d}\tilde{\rho}
    \Bigg].
    \label{eq:Laplace_Corner}
\end{multline}

\subsection{Cubic Gap Model}
\label{Cubic_Model}
The gap potential in Eq.~\eqref{eq31} may be approximated by higher-order polynomials. The admissible polynomial forms are restricted by the three potential conditions in Eq.~\eqref{eq19} and the endpoint electric-field symmetry condition $\vec{E}(0)=\vec{E}(1)$ in Eq.~\eqref{eq20}. For a quadratic polynomial, the potential conditions eliminate the quadratic term, reducing it to a linear form. For a quartic polynomial, imposing all four conditions eliminates the fourth-order term, leaving a cubic form. In the cubic case, the field-symmetry condition is automatically satisfied by the three potential conditions, leaving one adjustable coefficient. Thus, the cubic model provides the simplest nonlinear polynomial approximation that retains a free parameter to describe the variation of the gap potential. This coefficient is determined by fitting the model to the corner-effect gap potential. The resulting cubic approximation is written as
\begin{equation}
    V_g^i\!\left(\frac{\tilde{\xi}_i}{\tilde{w}_i}\right)
    =
    \sum_{j=0}^{3}
    \lambda_j
    \left(
    \frac{\tilde{\xi}_i}{\tilde{w}_i}
    \right)^j .
    \label{eq35}
\end{equation}
Applying the three potential conditions in Eq.~\eqref{eq19} to the cubic gap potential in Eq.~\eqref{eq35} gives $\lambda_0=V_i$, $\lambda_1=\Delta V_i+\lambda_3/2$, and $\lambda_2=-3\lambda_3/2$, where $\Delta V_i=V_{i+1}-V_i$. These relations also ensure that the endpoint field-symmetry condition in Eq.~\eqref{eq20} is satisfied. The cubic coefficient $\lambda_3$ remains a free parameter, whose value is determined by fitting the cubic model to the gap potential in Eq.~\eqref{eq31}. This fitting yields $\lambda_3=(19/22)\Delta V_i$. Substituting the resulting coefficients into Eq.~\eqref{eq35} gives the explicit form
\begin{multline}
    V_g^{(i)}\!\left(\frac{\tilde{\xi}_i}{\tilde{w}_i}\right)
    =
    V_i
    + \Delta V_i
    \left[
    \frac{63}{44}
    \left(
    \frac{\tilde{\xi}_i}{\tilde{w}_i}
    \right)
    - \frac{57}{44}
    \left(
    \frac{\tilde{\xi}_i}{\tilde{w}_i}
    \right)^2
    \right.\\
    \left.
    + \frac{19}{22}
    \left(
    \frac{\tilde{\xi}_i}{\tilde{w}_i}
    \right)^3
    \right].
    \label{eq40}
\end{multline}
Substituting the cubic gap potential of Eq.~\eqref{eq40} into Eq.~\eqref{eq:Lin_rad_gap_pot}, and applying the boundary conditions specified in Eqs.~\eqref{eq:electrode_bound_with_lin_gap}–\eqref{eq:gap_bound_with_lin_gap1}, together with the corresponding definition of $\phi_i^{g(2)}(\tilde{\rho},0)$ given by
\begin{equation}
    \phi_i^{g(2)}(\tilde{\rho}, 0) 
    =
    \begin{cases}
    0, & \tilde{\rho} < \tilde{a}_{i}, \\[4pt]
    \displaystyle
    \phi_{ab}^{(3)},
    & \tilde{a}_{i} \le \tilde{\rho} \le \tilde{b}_i, \\[8pt]
    0, & \tilde{\rho} > \tilde{b}_{i},
    \end{cases}
    \label{eq41}
\end{equation}
with
\begin{equation} \nonumber
  \phi_{ab}^{(3)}=  \frac{63}{44}\left( \frac{\tilde{\rho} - \tilde{a}_i}{\tilde{w}_i}\right)
    -
    \frac{57}{44}\left( \frac{\tilde{\rho} - \tilde{a}_i}{\tilde{w}_i}\right)^{2}
    \\
    +
    \frac{19}{22}\left( \frac{\tilde{\rho} - \tilde{a}_i}{\tilde{w}_i}\right)^{3},
\end{equation}
into the general solution of Eq.~\eqref{eq:Pot_General_Sol} and evaluating the resulting integral yields the axial potential
\begin{equation}
    V(0, \tilde{z}) =
    \sum_{i=1}^{N}
    \Delta {V}_i\,
    \Phi_i^{(3)}(\tilde{z}; \tilde{w}_i, \tilde{\rho}_c, \tilde{z}_c),
    \label{eq:Vz_Cubic}
\end{equation}
where the dimensionless potentials are given by 
\begin{widetext}
\begin{equation}
    \begin{split}
    \Phi_i^{(3)}(\tilde{z}; \tilde{w}_i, \tilde{\rho}_c, \tilde{z}_c)
    &=
    \sum_{n=1}^{\infty}
    \frac{2}{\tilde{\rho}_c^{2} J_{1}^{2}(\alpha_{0n})}
    \frac{\sinh\!\big(k_n(\tilde{z}-\tilde{z}_c)\big)}{\sinh(k_n \tilde{z}_c)} \\
    &\quad \times
    \Bigg[
    \frac{63}{44}
    \frac{1}{\tilde{w}_i}
    \Bigg\{
    \frac{\tilde{a}_i}{k_n^{2}} J_0(k_n \tilde{a}_i)
    - \frac{\tilde{b}_i}{k_n^{2}} J_0(k_n \tilde{b}_i)
    + \frac{1}{k_n^{2}}
    \int_{\tilde{a}_i}^{\tilde{b}_i}
    J_0(k_n \tilde{\rho})\, d\tilde{\rho}
    \Bigg\} \\[6pt]
    &\qquad
    +
    \frac{57}{44}
    \frac{1}{\tilde{w}_i^{2}}
    \Bigg\{
    \frac{4\tilde{a}_i}{k_n^{3}} J_1(k_n \tilde{a}_i)
    + \frac{4\tilde{b}_i}{k_n^{3}} J_1(k_n \tilde{b}_i)
    + \frac{2\tilde{a}_i}{k_n^{2}}
    \int_{\tilde{a}_i}^{\tilde{b}_i}
    J_0(k_n \tilde{\rho})\, d\tilde{\rho}
    \Bigg\} \\[6pt]
    &\qquad
    +
    \frac{19}{22}
    \frac{1}{\tilde{w}_i^{3}}
    \Bigg\{
    \frac{3\tilde{a}_i^{2}}{k_n^{3}} J_1(k_n \tilde{a}_i)
    - \frac{3\tilde{b}_i^{2}}{k_n^{3}} J_1(k_n \tilde{b}_i)
    + \frac{9\tilde{b}_i}{k_n^{4}} J_0(k_n \tilde{b}_i)
    - \frac{9\tilde{a}_i}{k_n^{4}} J_0(k_n \tilde{a}_i) \\
    &\qquad\qquad
    +
    \left(
    \frac{3\tilde{a}_i^{2}}{k_n^{2}}
    - \frac{9}{k_n^{4}}
    \right)
    \int_{\tilde{a}_i}^{\tilde{b}_i}
    J_0(k_n \tilde{\rho})\, d\tilde{\rho}
    \Bigg\}
    \Bigg] .
    \end{split}
    \label{eq:Laplace_Cubic}
\end{equation}
\end{widetext}
The consequences for the overall trap potential resulting from the gap contributions according to the various gap models (A)-(D) will enter the optimization process and will be discussed in detail in Sec.~\ref{effect}.

\section{Description of the Anharmonic Trap Potential}
\label{five}
\subsection{Series Expansion of the Potential}
\label{Exp_trap_pot}
The dynamics of the confined electron is governed entirely by the local behavior of the potential $V(\tilde{\rho},\tilde{z})$ in the vicinity of the trap center at $\tilde{z}_0$. In the otherwise charge-free trapping region, the electrostatic potential satisfies Laplace’s equation, and cylindrical symmetry implies that the full three-dimensional potential is uniquely determined by its axial values. It is therefore sufficient to consider the axial potential $\Phi(0,\tilde{z})$. To extract the trap properties in a systematic manner, the axial potential can be expanded in a Taylor series about the equilibrium position $\tilde{z}_0$. In this expansion, the quadratic term determines the axial oscillation frequency, while higher-order terms quantify anharmonicities of the confinement.

For the axial potential $\phi_i(0,\tilde{z})$ produced by a single electrode, this results in 
\begin{equation}
    \phi_i(0,\tilde{z})
    =
    \frac{1}{2}
    \sum_{k=0}^{\infty}
    C_{ki}(\tilde{z}-\tilde{z}_0)^k ,
    \label{eq45}
\end{equation}
where the expansion coefficients $C_{ki}$ are defined by
\begin{equation}
    C_{ki}
    =
    \frac{2}{k!}
    \left.
    \frac{\partial^{k}\phi_i(0,\tilde{z})}{\partial \tilde{z}^{k}}
    \right|_{\tilde{z}=\tilde{z}_0},
    \label{eq:Cki}
\end{equation}
where $k$ denotes the order of the derivative. 

The potentials evaluated in Eqs.~\eqref{eq:Laplace_Sol_Gapless},~\eqref{eq:Laplace_Lin}, \eqref{eq:Laplace_Corner}, and~\eqref{eq:Laplace_Cubic}, corresponding to the different gap-potential models (A)-(D), are analytic functions of the dimensionless trap geometry parameters $\{\tilde{\rho}_i, \tilde{w}_i, \tilde{\rho}_c, \tilde{z}_c\}$. Consequently, the coefficients $C_{ki}$ are likewise analytic functions of these geometry parameters, as well as of the equilibrium position $\tilde{z}_0$.

Similarly, the full axial potential of the trap can be expanded by invoking the principle of linear superposition together with Eq.~\eqref{eq45}. Introducing the dimensionless electrode voltages $\tilde{V}_i = V_i / V_0$, the total axial potential along the symmetry axis can be expanded as follows 
\begin{equation}
    V(0,\tilde{z})
    =
    V_0 \sum_{i=1}^{N} \tilde{V}_i \, \phi_i(0,\tilde{z})
    =
    \frac{V_0}{2}
    \sum_{k=0}^{\infty}
    C_k (\tilde{z}-\tilde{z}_0)^k,
    \label{eq:axial_pot_Ck}
\end{equation}
where the potential expansion coefficients $C_k$ of the total axial potential are related to the single-electrode coefficients through
\begin{equation}
    C_k
    =
    \sum_{i=1}^{N}
    C_{ki}\, \tilde{V}_i .
    \label{eq48}
\end{equation}
This equation forms the basis for optimizing the planar-trap parameters. The lack of mirror symmetry about the axial direction in planar traps leads to odd expansion coefficients generally being non-zero. 

The coefficient $C_{0}$ represents a constant potential offset and does not affect particle motion. The coefficient $C_{1}$ corresponds to a uniform axial electric field and shifts the equilibrium position from $\tilde z_0$. The coefficient $C_{2}$ is the desired quadrupole term responsible for axial confinement and primarily determines the axial oscillation frequency. The coefficient $C_{3}$ is the leading odd anharmonic term and can produce amplitude-dependent shifts in the axial oscillation frequency, whereas $C_{4}$ is the leading even anharmonic correction and produces symmetric distortions of the trapping potential. Higher-order coefficients $C_{5},C_{6},\ldots$ describe progressively higher multipole contributions whose effects are increasingly suppressed near the trap center.

Consider an electron executing small-amplitude axial oscillations about the equilibrium position $\tilde z=\tilde z_0$, with oscillation amplitude $\tilde A$. For $\tilde z_0$ to be an equilibrium position, $C_1$ must vanish; otherwise, the resulting axial electric field would displace the equilibrium. In the limit of vanishing amplitude, the potential energy is harmonic, $\tfrac{1}{2}m\omega_z^2(z-z_0)^2$. Comparing the quadratic term in Eq.~\eqref{eq:axial_pot_Ck}, multiplied by the electron charge $e$, with this harmonic form gives
\begin{equation}
    \omega_z=\sqrt{\frac{eV_0\,C_2}{m\rho_1^2}}.
    \label{eq:axial_freq2}
\end{equation}
This reduces to the ideal result of Eq.~\eqref{eq:axial_freq1} when the quadratic curvature is normalized by setting $C_2=1$. The resulting deviations from harmonicity are then characterized by the higher-order coefficients $C_3,C_4,\ldots$. The amplitude dependence of the axial oscillation frequency arising from these coefficients is considered next.

\subsection{Amplitude-Dependent Axial Frequency due to Anharmonic Potential}
\label{sec:anharmonic_frequency_link}

As discussed above, anharmonic terms in the axial potential, i.e., terms of order higher than $(\tilde z-\tilde z_0)^2$, introduce an amplitude dependence of the axial oscillation frequency $\omega_z$. 
Following the standard treatment of nonlinear oscillations using the Lindstedt--Poincar\'e method~\cite{mickens1981}, the frequency of this fundamental component can be expressed as a power series in the dimensionless axial oscillation amplitude $\tilde{A}$:
\begin{equation}
    \omega_z(\tilde{A})
    =
    \omega_z
    \left(
    1+\sum_{k=2}^{\infty}a_k\tilde{A}^k
    \right),
    \label{eq:freq_exp}
\end{equation}
where $\omega_z$ is the oscillation frequency in the limit of vanishing oscillation amplitude. From Eq.~\eqref{eq:freq_exp}, the relative frequency shift expressed in terms of the coefficients $a_k$ is given by
\begin{multline}
    \frac{\Delta\omega_z}{\omega_z}
    =
    \frac{\omega_z(\tilde{A})-\omega_z}
    {\omega_z}
    \\
    =
    a_2\tilde{A}^{2}
    +a_3\tilde{A}^{3}
    +a_4\tilde{A}^{4}
    +a_5\tilde{A}^{5}
    +\cdots .
    \label{eq:freq_shift}
\end{multline}
The coefficients $a_k$ are amplitude-expansion coefficients that depend uniquely on the above potential-expansion coefficients $C_k$ as defined in Eq.~\eqref{eq48}. 
Our analytic calculation of the coefficients $a_k$ as a function of the coefficients $C_k$ exactly reproduce the corresponding results given in \cite{goldman2010penning}, hence they are not reproduced here but we refer to equations (25) of \cite{goldman2010penning}. 
Consequently, non-zero values of $a_2, a_3, a_4, \ldots$ result in oscillation frequency shifts and spreads that depend on the oscillation amplitude.

\subsection{Thermal Width of the Axial Frequency Distribution}

In thermal equilibrium at temperature $T_z$, the probability that the axial electron energy lies between $E$ and $E+dE$ is given by the Boltzmann distribution
\begin{equation}
    P(E)\,dE=
    \frac{1}{k_BT_z}
    \exp\!\left(
    -\frac{E}{k_BT_z}
    \right)dE,
\end{equation}
where $E$ is related to the axial oscillation amplitude $A$ by
\begin{equation}
    E=\frac{1}{2}m\omega_z^2A^2.
    \label{eq:energy_amp}
\end{equation}
The probability density of the oscillation amplitude is obtained by a change of variables from $E$ to $A$, such that $P(A)\,dA=P(E)\,dE$,
leading to
\begin{equation}
    P(A)\,dA
    =
    \frac{m\omega_z^2}{k_BT_z}
    A
    \exp\!\left(
    -\frac{m\omega_z^2A^2}{2k_BT_z}
    \right)dA,
\end{equation}
which is a Rayleigh distribution \cite{Papoulis2002}. Expressed in terms of the scaled oscillation amplitude, $\tilde A \equiv A/\rho_1$, the probability density becomes
\begin{equation}
    P(\tilde A)
    =
    \frac{\tilde A}{\tilde A_T^2}
    \exp\!\left(
    -\frac{\tilde A^2}{2\tilde A_T^2}
    \right),
    \label{eq:rayleigh_amp}
\end{equation}
where
\begin{equation}
    \tilde A_T
    =
    \sqrt{
    \frac{k_BT_z}
    {m\omega_z^2\rho_1^2}
    }
    \label{eq:rayleigh_scale}
\end{equation}
is the Rayleigh scale parameter, determined by the equipartition theorem for a classical harmonic oscillator \cite{Reif1965,Pathria2011}.
Consequently, the thermally averaged frequency shift (i.e. the shift of the center of the frequency distribution with respect to the $T=0$ limit) is
\begin{equation}
    \Delta f_z
    =
    f_z
    \left\langle
    \frac{\Delta\omega_z}{\omega_z}
    \right\rangle,
    \label{eq:thermal_shift}
\end{equation}
while the thermal width of the frequency distribution is
\begin{equation}
    \sigma_f
    =
    f_z
    \sqrt{
    \left\langle
    \left(
    \frac{\Delta\omega_z}{\omega_z}
    \right)^2
    \right\rangle
    -
    \left\langle
    \frac{\Delta\omega_z}{\omega_z}
    \right\rangle^2
    },
    \label{eq:thermal_spread}
\end{equation}
where the averages are taken over the thermal Rayleigh amplitude distribution given by Eq.~\eqref{eq:rayleigh_amp}. The higher-order coefficients $a_k$ therefore contribute not only to the mean frequency shift but also to the thermal broadening of the axial resonance \cite{brown1986}.

\section{Harmonic Optimization of a Three-Gap Planar Trap}
\label{Trap_Tunability}
To minimize amplitude-dependent frequency shifts $\Delta f_{z} (a_k)$ and thermal broadening $\sigma_f(a_k)$, the trap geometry and electrode voltages are optimized such that the coefficients $a_k$ vanish. We focus on the particularly promising configuration 
\begin{equation}
\begin{aligned}
    C_1&=0,\qquad C_2=1,\qquad C_3=C_4=C_6=0,
    \end{aligned}
    \label{eq:Opt_path_1}
\end{equation}
for which the leading amplitude-expansion coefficients vanish, $a_2=a_3=a_4=a_5=0$. 

A three-gap planar trap has six independent scaled parameters, namely $(\tilde{\rho}_2,\tilde{\rho}_3,\tilde{V}_1,\tilde{V}_2,\tilde{V}_3, \tilde{z}_0)$. The constraints $C_1=0$ and $C_2=1$ define the equilibrium position and set the axial curvature, reducing the number of independent parameters to four. Since the amplitude coefficients $a_k$ depend only on the higher-order potential coefficients $C_{k\geq3}$, the leading amplitude dependence is removed by requiring $a_2=0$, which is satisfied for $C_3=C_4=0$. These conditions leave two independent geometric parameters, corresponding to the two scaled electrode radii. One additional constraint, such as $C_5=0$ or $C_6=0$, can be imposed to further reduce the remaining freedom. However, imposing $C_5=0$ instead of $C_6=0$ would leave a nonzero value of $a_4$, whereas imposing $C_6=0$ allows the amplitude coefficients to vanish through $a_5$.

Thus, unlike the two-gap planar trap, the three-gap geometry provides sufficient degrees of freedom to simultaneously satisfy the equilibrium condition $C_1=0$, curvature normalization $C_2=1$, and higher-order anharmonicity constraints $C_{k\geq3}$. This enables planar Penning traps with fifth-order suppression of amplitude-dependent frequency shifts and substantially improved harmonicity.
The task is to determine the corresponding electrode voltages and geometry parameters $(\tilde{\rho}_2,\tilde{\rho}_3,\tilde{V}_1,\tilde{V}_2,
    \tilde{V}_3,\tilde{z}_0)$
for different boundary conditions and gap-potential models.

In the following discussions, two approaches to optimization of the potential are presented: one is via optimisation of the voltages only. This is called ``in-operando'' since it can be used to tune the trap during an experiment. The other is via optimisation of the voltages and the trap geometry. This is called ``ex-operando'' since it requires changes to the hardware, but generally introduces more freedom for optimisation.

\section{Effect of finite boundaries}
\label{boun}
To investigate the effect of finite electrostatic boundaries on the optimized trap parameters, we first consider the zero-gap limit and compare the optimized solutions obtained for infinite and finite boundaries.

\subsection{Optimized trap parameters for \textbf{infinite} boundaries under the zero-gap approximation}
\label{infbound}

\begin{figure}[t]
    \centering
    \includegraphics[width=0.9\columnwidth]{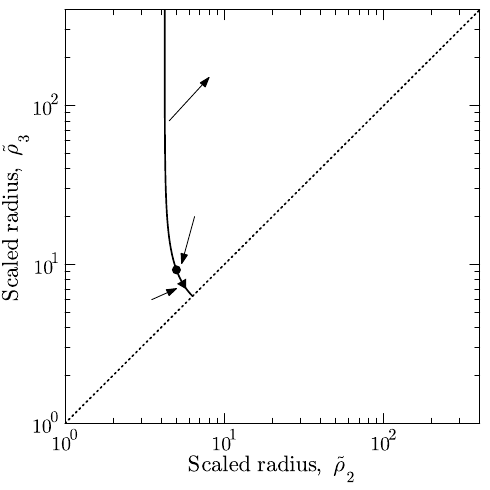}
    \caption{Optimized geometry for a three-gap planar trap with $\tilde{\rho}_1=1$, $\tilde{w}_i=0$, and $\tilde{\rho}_c=\tilde{z}_c=\infty$, satisfying $a_2=a_3=a_4=a_5=0$. The circle marks the parameter set used for detailed analysis, while the triangle indicates the optimized parameters used in \cite{goldman2010penning}.}
    \label{fig:R3s_vs_R2s}
\end{figure}

We begin with the simplest case, assuming infinitesimal gaps between the electrodes and external boundaries extending to infinity. With $\tilde{\rho}_1=1$ by definition, the optimized scaled voltages are obtained for different geometric configurations $(\tilde{\rho}_2,\tilde{\rho}_3)$ using the analytical axial potential derived under the infinitesimal-gap and infinite-boundary approximations, as presented in Sec.~\ref{Infinitesimal_Gaps}. Imposing the constraints in Eq.~\eqref{eq:Opt_path_1} yields the corresponding optimized solutions and defines the admissible range of geometries shown in Fig.~\ref{fig:R3s_vs_R2s}.

The allowed solutions lie above the line $\tilde{\rho}_2 = \tilde{\rho}_3$, while the region below is nonphysical due to the violation of electrode ordering. The admissible configurations form a narrow band, indicating a strong geometric correlation between $\tilde{\rho}_2$ and $\tilde{\rho}_3$. In particular, $\tilde{\rho}_3$ decreases rapidly with increasing $\tilde{\rho}_2$, reflecting the need to maintain the optimized harmonic conditions of Eq.~\eqref{eq:Opt_path_1}.

To analyze the various properties of the trap, it is necessary to focus on a specific optimized geometrical configuration. Although, in principle, any point along the admissible solution curve shown in Fig.~\ref{fig:R3s_vs_R2s} can yield a highly harmonic planar trap, practical considerations favor configurations in which the required electrode voltages remain within a moderate range. The triangle in Fig.~\ref{fig:R3s_vs_R2s} represents the sample trap geometry studied in \cite{goldman2010penning}. Our selected configuration requires a lower maximum electrode voltage, making it more practical to implement using available high-stability DC power supplies. Accordingly, we select the configuration indicated by the solid circle in Fig.~\ref{fig:R3s_vs_R2s}, which we refer to as the \emph{sample trap}, defined by
\begin{equation}
\tilde{\rho}_i = \{1.0,\, 5.0,\, \tilde{\rho}_3\}.
\end{equation}
Here, the optimized value of $\tilde{\rho}_3$ depends on the external boundary conditions as well as on the gap potential model used to satisfy the imposed constraints in Eq.~\eqref{eq:Opt_path_1} for a given $\tilde{\rho}_1$ and $\tilde{\rho}_2$. For infinite external boundaries and the zero-gap approximation, the optimized value of $\tilde{\rho}_3$ is $9.2408$. The optimized radius $\tilde{\rho}_3$ and the corresponding potentials are reported here up to four decimal places; however, the analytical calculations are performed with a precision of up to 12 decimal places, achieving a tolerance of order $10^{-12}$ in the coefficients $C_k$.
The optimized values satisfying the constraints in Eq.~\eqref{eq:Opt_path_1} correspond to the equilibrium position $\tilde{z}_0 = 1.4342$.

Further, to evaluate the trap properties, two of the three parameters $\{f_z, \rho_1, V_0\}$ must be specified. In the present work, we fix the axial frequency at $f_z = 4\,\mathrm{GHz}$, which corresponds to a required potential curvature of $-3591.35\,\mathrm{MV/m^2}$ for electron confinement. For a characteristic length scale of $\rho_1 = 10\,\mu\mathrm{m}$, this yields a potential depth of $V_0 = -0.3591\,\mathrm{V}$, corresponding to the solid circle in Fig.~\ref{fig:curvature_fz}(b).

\begin{figure}[ht]
    \centering
    \includegraphics[width=\columnwidth]{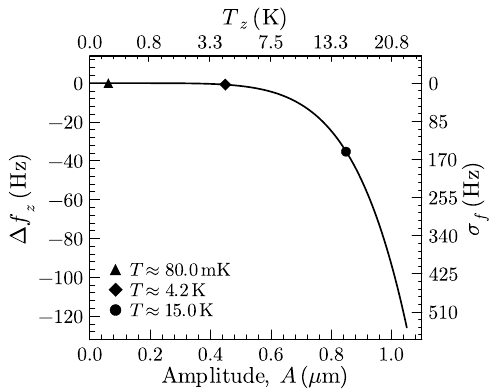}
    \caption{Amplitude dependence of the axial frequency shift $\Delta f_z$ (left-hand scale) and the thermal frequency spread $\sigma_f$ (right-hand scale) for the sample trap with a nominal axial frequency $f_z=4\,\mathrm{GHz}$, computed using the optimized parameters listed in Table~\ref{tab:optimized_parameters_zero_gap_inf_finite_bound}. 
    }
    \label{fig:thermal_freq_spreads_gapless_inf_bound}
\end{figure}
The optimized parameters of the sample trap are presented in Table~\ref{tab:optimized_parameters_zero_gap_inf_finite_bound} and contrasted with the corresponding values for finite boundaries as we will discuss below. 

The thermal frequency shift $\Delta f_z$ (relative to $T=0$) and the associated frequency spread $\sigma_f$ at finite temperature are calculated using Eqs.~\eqref{eq:thermal_shift} and \eqref{eq:thermal_spread}, together with Eq.~\eqref{eq:rayleigh_scale}, using the optimized coefficients $a_k$. Throughout this work, unless stated otherwise, these quantities are evaluated using the optimized expansion coefficients up to $a_5$, for which both the thermal frequency shift and frequency spread vanish identically, i.e., $\Delta f_z=0$ and $\sigma_f=0$. We also evaluate the effect of the finite $a_6$ coefficient, which represents the leading contribution beyond this optimized expansion, as shown in Fig.~\ref{fig:thermal_freq_spreads_gapless_inf_bound} and Table~\ref{tab:optimized_parameters_zero_gap_inf_finite_bound}. Enforcing $a_6=0$ requires additional control over the otherwise unconstrained coefficients $C_5$ and $C_8$, thereby requiring an additional degree of freedom available only in a four-gap planar trap configuration. Including the $a_6$ contribution yields $|\Delta f_z| \simeq 0.656\,\mathrm{Hz}$ and $\sigma_f \simeq 2.860\,\mathrm{Hz}$ at $T_z=4\,\mathrm{K}$. These values correspond to approximately $1.6\times10^{-10}$ and $7.2\times10^{-10}$, respectively, of the axial frequency $f_z=4\,\mathrm{GHz}$. 

The three data points highlighted in Fig.~\ref{fig:thermal_freq_spreads_gapless_inf_bound} correspond to temperatures of $80\,\mathrm{mK}$, $4.2\,\mathrm{K}$, and $15\,\mathrm{K}$, representing common experimentally relevant operating temperatures for a $^3$He/$^4$He dilution refrigerator, a liquid-helium (LHe) cryostat, and a common closed-cycle cryo-cooler, respectively. The corresponding absolute thermal frequency shifts and frequency spreads, $(|\Delta f_z|,\sigma_f)$, are $(6.0\,\mu\mathrm{Hz},26.0\,\mu\mathrm{Hz})$, $(0.76\,\mathrm{Hz},3.3\,\mathrm{Hz})$, and $(35.3\,\mathrm{Hz},152.7\,\mathrm{Hz})$, respectively.

\subsection{Optimized trap parameters for \textbf{finite} boundaries under the zero-gap approximation}
\label{finbound}

In practical implementations, external boundaries are necessarily finite and can significantly influence trap performance, particularly when their dimensions are comparable to the trap size. To model this effect, we consider a grounded cylindrical enclosure with radial and axial dimensions $\rho_c$ and $z_c$, respectively. Two complementary approaches are used to assess and compensate for the effect of these boundaries.

\subsubsection{First Approach: Fixed Geometry, Re-Optimized Voltages}
\begin{figure}[!h]
    \centering
    \includegraphics[width=0.9\columnwidth]{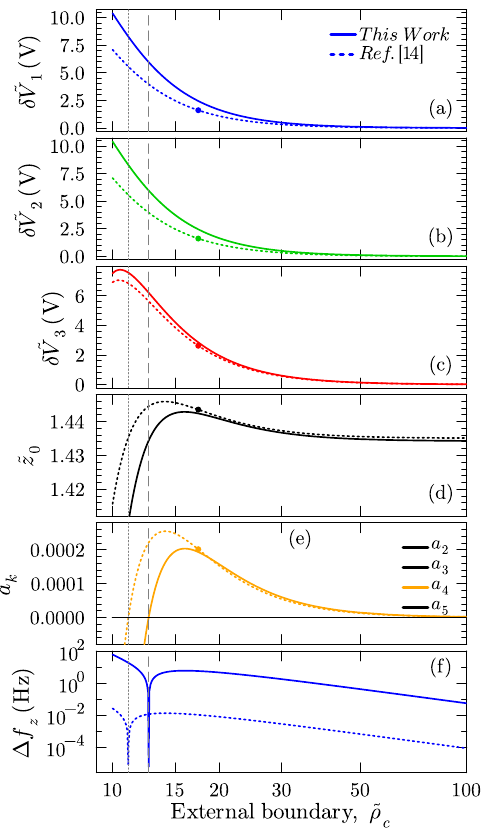}
    \caption{Effect of finite external boundaries on the optimized trap under the zero-gap approximation. Solid and dotted curves correspond to the sample-trap geometries $\{1.0,\,5.0,\,9.2408\}$ with $\tilde{z}_c=3\tilde{\rho}_c$ and $\{1.0,\,5.5,\,7.5426\}$ with $\tilde{z}_c=2.4\tilde{\rho}_c$, respectively. The panels show (a)–(c) the voltage shifts required to maintain $C_1=C_3=C_4=0$ and $C_2=1$, (d) the equilibrium position $\tilde{z}_0$, (e) the amplitude coefficients $a_k$, and (f) the absolute thermal frequency shift $\Delta f_z$ at $4\,\mathrm{K}$. Vertical lines mark the finite-boundary configurations for which $a_4=0$: $\tilde{\rho}_c\approx12.6452$ (solid curves) and $\tilde{\rho}_c\approx11.0799$ (dotted curves). At these points, $C_6=0$ and $a_2=a_3=a_4=a_5=0$. Solid circles on the dotted curves indicate Solution~III of Table~VI in Ref.~\cite{goldman2010penning}, corresponding to $\tilde{\rho}_c\approx17.46$ and $\tilde{z}_c\approx41.91$ for $\rho_1=1.0909\,\mathrm{mm}$.}
    \label{fig:Effect of external boundaries}
\end{figure}

In the first approach, the sample-trap geometry $\{1.0,\,5.0,\,9.2408\}$ is kept fixed, while the electrode potentials are adjusted to compensate for the anharmonicity introduced by finite boundaries. The radial boundary $\tilde{\rho}_c$ is varied from $10.0$, slightly larger than $\tilde{\rho}_3=9.2408$, to $\sim30\,\tilde{\rho}_3$, while the axial boundary is chosen as $\tilde{z}_c=3\tilde{\rho}_c$; more generally, it is sufficient that $\tilde{z}_c\gg\tilde{z}_0$. For the fixed geometry, the three independent electrode voltages allow the constraints $C_1=0$, $C_2=1$, and $C_3=C_4=0$ to be imposed simultaneously. Because $\tilde{\rho}_3$ is fixed, however, the additional condition $C_6=0$, which together with $C_3=C_4=0$ ensures $a_2=a_3=a_4=a_5=0$, cannot generally be satisfied. The voltage corrections relative to the optimized infinite-boundary values are obtained from the finite-boundary solution of Laplace's equation under the zero-gap approximation, Eq.~\eqref{eq:Laplace_Sol_Gapless}, together with the corresponding axial potential in Eq.~\eqref{eq:Vz_gapless}. The solid curves in Fig.~\ref{fig:Effect of external boundaries} show the resulting voltage shifts, amplitude coefficients $a_k$, and absolute thermal frequency shift $\Delta f_z$ at $4\,\mathrm{K}$ as functions of $\tilde{\rho}_c$, with $\Delta f_z$ evaluated by retaining terms through $a_5$ for the present work. The dotted curves show the corresponding results for the sample-trap geometry $\{1.0,\,5.5,\,7.5426\}$ with $\tilde{z}_c=2.4\tilde{\rho}_c$, as considered in Ref.~\cite{goldman2010penning}. Both sets of results are presented up to $\tilde{\rho}_c=100$ to highlight features at smaller boundary dimensions.

For small $\tilde{\rho}_c$, finite boundaries strongly perturb the trap parameters relative to the infinite-boundary case, requiring substantial corrections to the electrode voltages [Fig.~\ref{fig:Effect of external boundaries}(a)–(c)]. As the boundary dimensions increase, these corrections rapidly decrease, and the trap parameters approach their optimized infinite-boundary values. The amplitude coefficients $a_k$ in Fig.~\ref{fig:Effect of external boundaries}(e) show that, although the constraints $C_1=0$, $C_2=1$, $C_3=0$, and $C_4=0$ are enforced throughout, the residual finite value of $C_6$ gives rise to a nonzero $a_4$ and hence residual amplitude dependence of the axial frequency. This residual anharmonicity is reflected in the nonzero thermal frequency shift shown in Fig.~\ref{fig:Effect of external boundaries}(f).

A notable feature is that $C_6$ has two zeros: one at the finite-boundary location $\tilde{\rho}_c=12.6452$, indicated by the vertical dashed line for the solid curves, and the other in the limit $\tilde{\rho}_c\rightarrow\infty$. At both boundary configurations, the constraint set defined by Eq.~\eqref{eq:Opt_path_1} is satisfied, yielding $a_2=a_3=a_4=a_5=0$. At the finite-boundary harmonic point, the amplitude coefficients through $a_5$ and the thermal frequency shift $\Delta f_z$ vanish. Within the $a_5$-order treatment, the thermal frequency spread $\sigma_f$ likewise vanishes (not plotted here), indicating suppression of the axial-frequency amplitude dependence through fifth order and realization of the optimized harmonic configuration. Away from these special boundary configurations, finite values of $C_6$ generate residual anharmonicity and hence nonzero thermal frequency shifts and spreads. The equilibrium position $\tilde{z}_0$ [Fig.~\ref{fig:Effect of external boundaries}(d)] exhibits similar behavior: significant deviations occur for small boundary dimensions, whereas it approaches its optimized value of $\tilde{z}_0=1.4342$ near $\tilde{\rho}_c=12.6452$ and in the infinite-boundary limit. The recovery of the optimized voltage shifts, amplitude coefficients, and thermal frequency characteristics provides a consistency check on the finite-boundary formulation.

For comparison, the dotted curves in Fig.~\ref{fig:Effect of external boundaries} represent the results for the sample-trap geometry and boundary aspect ratio adopted in Ref.~\cite{goldman2010penning}. These curves exhibit similar trends throughout the investigated range of $\tilde{\rho}_c$, with the harmonic condition occurring at $\tilde{\rho}_c\approx11.0799$, as indicated by the vertical dotted line, and in the infinite-boundary limit, where $C_6=0$ and $a_2=a_3=a_4=a_5=0$. The solid circles on the dotted curves mark the results for Solution III in Table~VI of Ref.~\cite{goldman2010penning}, reproduced for $\tilde{\rho}_c\approx17.46$ and $\tilde{z}_c\approx41.91$ ($\tilde{z}_c\approx2.4\tilde{\rho}_c$) with $\rho_1=1.0909\,\mathrm{mm}$. Differences between the two sets of results (solid and dotted curves) arise from the different axial boundary dimensions, with $\tilde{z}_c=3\tilde{\rho}_c$ in the present work. As $\tilde{\rho}_c\to\infty$, the results agree closely, except for $\tilde{z}_0$, owing to the different sample trap geometries. Thus, the differences reflect the optimization conditions and solution branches considered. To achieve the desired harmonicity under the same optimization conditions defined by Eq.~\eqref{eq:Opt_path_1}, $\tilde{\rho}_3$ must also be adjusted along with the electrode voltages for the given external boundary dimensions.

\subsubsection{Second Approach: Re-Optimized Geometry and Re-Optimized Voltages}
An alternative approach is to re-optimize the sample-trap geometry and the corresponding electrode potentials to compensate for the effects of finite external boundaries. 
Using the axial potential in Eq.~\eqref{eq:Vz_gapless}, together with the solution of Laplace's equation for finite boundaries in Eq.~\eqref{eq:Laplace_Sol_Gapless} under the zero-gap approximation, the trap parameters are re-optimized for these boundary conditions. The optimized parameters of the sample trap for finite boundaries $\{\rho_c,\,z_c\}=\{250\,\mu\mathrm{m},\, 750\,\mu\mathrm{m}\}$ are listed in Table~\ref{tab:optimized_parameters_zero_gap_inf_finite_bound}, alongside those of the corresponding optimized configuration in the infinite-boundary limit. For $\rho_1=10,\mu\mathrm{m}$, the corresponding scaled external boundaries are $\{\tilde{\rho}_c,\,\tilde{z}_c\}=\{25,\,75\}$, which are considerably larger than the optimized value of $\tilde{\rho}_3$.
\begin{table}[t]
\centering
    \caption{Optimized parameters for the sample trap with infinite and finite external boundaries under the zero-gap approximation, corresponding to the optimized harmonic configuration defined by Eq.~\eqref{eq:Opt_path_1}.
\label{tab:optimized_parameters_zero_gap_inf_finite_bound}}
\begin{tabular}{lcc}
\hline\hline
\multicolumn{3}{c}{
\begin{tabular}{c}
Optimized for $V_0 = -0.3591\,\mathrm{V}$, $f_z = 4\,\mathrm{GHz}$, $\rho_1 = 10\,\mu\mathrm{m}$, \\
Scaled radii $\tilde{\rho}_i = \{1.0,\, 5.0,\, \tilde{\rho}_3\}$, \\
Imposed constraints $C_1 = 0$, $C_2 = 1$, and \\
$C_3 = C_4 = C_6 = 0 \Rightarrow a_2 = a_3 = a_4 = a_5 = 0$.
\end{tabular}
} \\
\hline
Parameter & \multicolumn{1}{c}{{Infinite} boundary} & \multicolumn{1}{c}{{Finite} boundary} \\
          & $(\rho_c,z_c)\to\infty$ & $(\rho_c,z_c)=(250,\,750)\,\mu\mathrm{m}$ \\
\hline
$\tilde{\rho}_3$ & 9.2408 & 9.1888 \\
$\tilde{V}_1$ & $-30.4163$ & $-29.4017$ \\
$\tilde{V}_2$ & $-31.0786$ & $-30.0639$ \\
$\tilde{V}_3$ & $-73.7132$ & $-72.7628$ \\
$\tilde{z}_0$ & $1.4342$ & $1.4342$ \\
$C_5$ & $-0.0107$ & $-0.0107$ \\
$C_8$ & $-0.0033$ & $-0.0033$\\
$a_6$ & $-0.0039$ & $-0.0039$ \\
$\Delta f_z~(\mathrm{Hz})~@~4\,\mathrm{K}$ & $0.656$ & $0.656$ \\
$\sigma_f~(\mathrm{Hz})~@~4\,\mathrm{K}$ & $2.860$ & $2.860$ \\
\hline\hline
\end{tabular}
\end{table}
Both configurations exhibit nearly identical trap properties, with the finite external boundaries primarily changing the optimized value of $\tilde{\rho}_3$ and the corresponding scaled electrode potentials. Although there is an overall shift in the axial potentials due to the different external boundaries, both configurations exhibit identical harmonic properties, as reflected in Table~\ref{tab:optimized_parameters_zero_gap_inf_finite_bound}. Since both configurations are optimized for the same harmonic configuration defined by Eq.~\eqref{eq:Opt_path_1}, they yield identical $C_k$ and the resulting $a_k$ coefficients. 

\section{Effect of finite electrode gaps}
\label{effect}
We discuss the axial potential of the trap with finite electrode gaps under finite external boundaries, using the gap-potential models (B)--(D) introduced in Sec.~\ref{gap_potentials}. Model (A) is not considered here, as it corresponds to the zero-gap case treated in the previous section. 
An exact analytical treatment that simultaneously accounts for finite gap width and depth is difficult; however, when the adjacent electrode widths are much larger than the gap width and the electrode height is large, the gap depth can be neglected, allowing the finite-gap effect to be described using approximate surface gap-potential models.


Three models are considered: the Linear (Sec.~\ref{Linear_Model}), Cubic (Sec.~\ref{Cubic_Model}), and Corner-Law (Sec.~\ref{Corner-Law_Model}) gap-potential models. The Linear model provides the simplest approximation by assuming a linear interpolation of the potential across the gap, whereas the Cubic model introduces higher-order spatial variation and thus represents the next step beyond the linear approximation. The Corner-Law model additionally accounts for field enhancement near electrode edges and corners, providing a more physically realistic description of the gap region. The Cubic model is particularly useful when a nonlinear description beyond the linear model is needed without the additional computational effort associated with the Corner-Law model.

As discussed in the previous section, two optimization strategies are possible: re-optimizing only the electrode potentials while keeping the geometry fixed, or re-optimizing both the geometrical parameter $\tilde{\rho}_3$ and the electrode potentials to satisfy the complete constraint set of Eq.~\eqref{eq:Opt_path_1}. 

\subsection{First Approach: Fixed Geometry, Re-Optimized Voltages}
In Ref.~\cite{goldman2010penning} it has been demonstrated that, for a scaled gap width of $\tilde{w}_i\approx0.046$ (corresponding to a physical gap width of $50\,\mu\mathrm{m}$ and $\rho_1=1.0909\,\mathrm{mm}$), adjusting the electrode potentials under the linear model can minimize the anharmonicity while maintaining the optimized trap geometry. We first examine this approach over the range $0\leq\tilde{w}_i\leq0.5$ by comparing our sample trap with infinite boundaries against the trap studied in Ref.~\cite{goldman2010penning}. We then consider finite external boundaries and apply the second approach using all three gap-potential models.
\begin{figure}[ht]
    \centering
    \includegraphics[width=0.9\columnwidth]{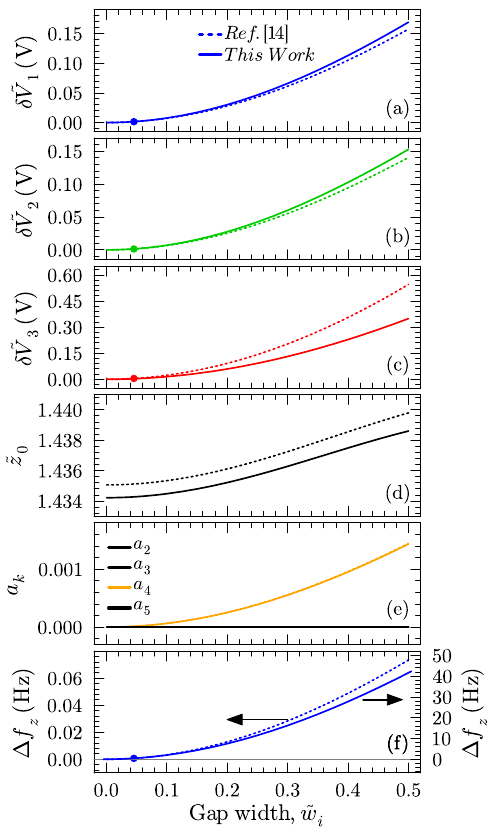}
        \caption{Effect of finite gap width on optimized traps with infinite external boundaries under the Linear gap-potential model. Solid and dotted curves correspond to the sample trap geometries $\{1.0,5.0,9.2408\}$ and $\{1.0,5.5,7.5426\}$, respectively, the latter corresponding to the geometry considered in Ref.~\cite{goldman2010penning}. Panels (a)--(c) show the voltage shifts required to maintain $C_1=C_3=C_4=0$ and $C_2=1$; panels (d)--(f) show the equilibrium position $\tilde{z}_0$, the amplitude coefficients $a_k$, and the absolute thermal frequency shift $\Delta f_z$ at $4\,\mathrm{K}$, respectively. The solid circles on the dotted curves mark the results for Solution III in Table~IV of Ref.~\cite{goldman2010penning}, corresponding to $\tilde{w}_i\approx0.046$. For both sample traps, the thermal frequency shift remains relatively small for $\tilde{w}_i\leq0.05$ but increases at larger gap widths. This increase is associated with a nonzero $C_6$, which gives rise to a finite $a_4$ and, consequently, residual anharmonicity.}
    \label{fig:Effect of finite gap widths with Linear gap model}
\end{figure}

The effect of finite gap width on the optimized trap parameters under the linear gap-potential model is shown in Fig.~\ref{fig:Effect of finite gap widths with Linear gap model}. The solid and dotted curves correspond to the sample-trap geometries considered in the present work and in Ref.~\cite{goldman2010penning}, respectively.  Panels (a)--(c) show the voltage shifts relative to the optimized electrode potentials of the corresponding zero-gap traps with infinite boundaries, listed in Table~\ref{tab:optimized_parameters_zero_gap_inf_finite_bound} for the present work and for Solution~III in Table~I of Ref.~\cite{goldman2010penning}. Panel (d) shows the equilibrium position $\tilde{z}_0$, panel (e) shows the amplitude coefficients $a_k$, and panel (f) shows the absolute thermal frequency shift $\Delta f_z$ at $4\,\mathrm{K}$, evaluated by retaining terms through $a_5$. The thermal frequency spread was also evaluated but is not shown, as it was not reported in Ref.~\cite{goldman2010penning}, and is therefore omitted to maintain a consistent comparison. The solid circles on the dotted curves indicate the results for Solution~III in Table~IV of Ref.~\cite{goldman2010penning} at $\tilde{w}_i\approx0.046$. At this gap width, the three voltage shifts closely match those obtained in the present work while satisfying $C_1=C_3=C_4=0$ and $C_2=1$.


The results show that the influence of finite gap width on the optimized parameters of the sample trap becomes increasingly pronounced as the gap width increases. Over the investigated range ($0\leq\tilde{w}_i\leq0.5$), the changes in the optimized trap parameters and the resulting thermal frequency shift remain relatively small for $\tilde{w}_i\leq0.05$. In this regime, the effect of finite gap width can be effectively compensated for by re-optimizing the electrode potentials alone, without changing the trap geometry. This is consistent with the results reported for Solution III in Table~IV of Ref.~\cite{goldman2010penning} at $\tilde{w}_i\approx0.046$. 

However, at larger gap widths, the deviations become more pronounced. The emergence of a finite $a_4$, associated with a nonzero $C_6$, indicates residual anharmonicity that cannot be eliminated by adjusting the electrode potentials alone while keeping the geometry fixed. Therefore, both the geometrical parameter $\tilde{\rho}_3$ and the electrode potentials must be re-optimized to satisfy the complete constraint set of Eq.~\eqref{eq:Opt_path_1}. Furthermore, the accuracy of this optimization depends on the realistic representation of the spatial variation of the potential across the electrode gaps. In particular, the Linear model assumes a constant electric field across each gap and therefore cannot capture the nonlinear potential variation and field enhancement near electrode edges. These effects become increasingly important as the gap width increases, motivating the second approach, in which both the trap geometry and electrode potentials are re-optimized using more realistic nonlinear gap-potential models.

\subsection{Second Approach: Re-Optimized Geometry and Re-Optimized Voltages}
Consequently, we apply the second approach to the sample trap with fixed external boundaries $\{\rho_c,\, z_c \}=\{250,\,750\}\,\mu\mathrm{m}$. The optimized trap parameters are evaluated over the gap-width range $0\leq\tilde{w}_i\leq0.5$ using the three gap-potential models (B)--(D), while enforcing the complete constraint set of Eq.~\eqref{eq:Opt_path_1}. 

\begin{figure}[t]
    \centering
    \includegraphics[width=0.9\columnwidth]{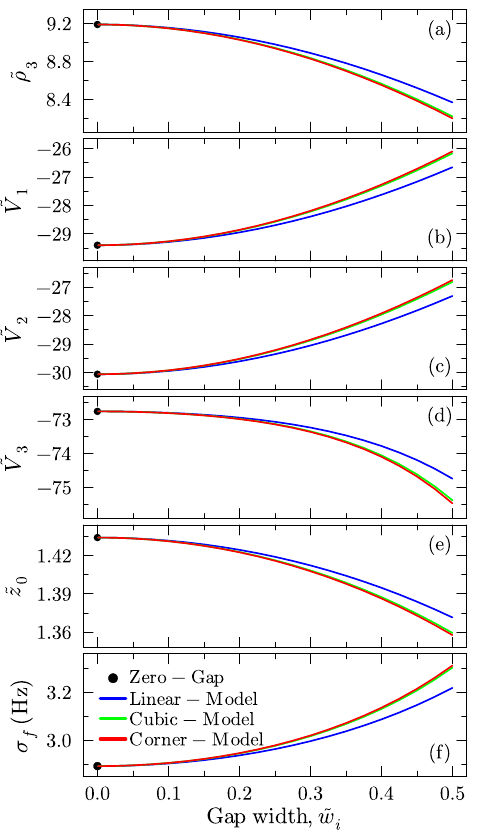}
    \caption{Effect of finite gap widths on the optimized trap parameters for fixed external boundaries $\{\rho_c,z_c\}=\{250,\,750\}\,\mu\mathrm{m}$, obtained using the Linear, Cubic, and Corner-Law gap-potential models under the constraints of Eq.~\eqref{eq:Opt_path_1}: (a) optimized $\tilde{\rho}_3$; (b)--(d) corresponding scaled electrode potentials $\tilde{V}_1$, $\tilde{V}_2$, and $\tilde{V}_3$; (e) equilibrium position $\tilde{z}_0$; and (f) thermal frequency spread $\sigma_f$ at $T_z=4\,\mathrm{K}$, calculated by retaining anharmonic coefficients through $a_6$. The zero-gap results (solid circles at $\tilde{w}_i=0$) are shown for reference.}
    \label{fig:Effect of finite gap widths with different gap models}
\end{figure}

Fig.~\ref{fig:Effect of finite gap widths with different gap models} shows the optimized $\tilde{\rho}_3$, the scaled electrode potentials $\tilde{V}_i$, the equilibrium position $\tilde{z}_0$, and the thermal frequency spread $\sigma_f$ at $4\,\mathrm{K}$. The corresponding zero-gap solution is included for reference. Both the thermal frequency shift and spread are evaluated by retaining terms through $a_6$, since their contributions vanish when the expansion is truncated at $a_5$.

As expected, all three gap-potential models recover the zero-gap solution as $\tilde{w}_i\rightarrow0$. As the gap width increases, finite-gap effects become progressively more significant, and the optimized parameters vary smoothly to account for the modified gap boundary conditions while maintaining the constraints of Eq.~\eqref{eq:Opt_path_1}. In particular, $\tilde{\rho}_3$ [Fig.~\ref{fig:Effect of finite gap widths with different gap models}(a)] and $\tilde{z}_0$ [Fig.~\ref{fig:Effect of finite gap widths with different gap models}(e)] decrease with increasing $\tilde{w}_i$, while $\tilde{V}_3$ [Fig.~\ref{fig:Effect of finite gap widths with different gap models}(d)] becomes progressively more negative and $\tilde{V}_1$ [Fig.~\ref{fig:Effect of finite gap widths with different gap models}(b)] and $\tilde{V}_2$ [Fig.~\ref{fig:Effect of finite gap widths with different gap models}(c)] become less negative. 

The Linear model predicts the smallest deviations from the zero-gap solution, whereas the Cubic and Corner-Law models exhibit very similar behavior for scaled gap widths up to $\tilde{w}_i\approx0.2$, beyond which their predictions begin to diverge noticeably. The deviations of both nonlinear models from the Linear model become more pronounced for $\tilde{w}_i\gtrsim0.1$, highlighting the importance of including nonlinear spatial variation in the gap potential even at moderate gap widths. This difference is important because re-optimizing the electrode geometry and voltages does not eliminate errors arising from an inadequate representation of the gap potential. 

The Linear model can satisfy the imposed harmonicity constraints within its assumed potential approximation, but this does not necessarily imply that it accurately represents the electrostatic potential of the actual finite-gap geometry. Consequently, its predictions for the optimized geometry, electrode voltages, and residual higher-order anharmonic contributions may deviate from those obtained using more realistic gap-potential descriptions.

The differences between the Cubic and Corner-Law models at larger gap widths arise from the singular behavior of the electric field near conducting edges. Electrostatic solutions of Laplace's equation exhibit a power-law dependence, $\Phi\propto\rho^\lambda$, where $\lambda$ depends on the wedge angle. For the re-entrant $270^\circ$ conducting corners relevant to the present geometry, $\Phi\propto\rho^{2/3}$ and consequently $E\propto\rho^{-1/3}$. The Cubic model, being a third-order polynomial approximation, cannot reproduce this edge singularity. Nevertheless, it provides a useful nonlinear approximation to the gap potential and closely reproduces the predictions of the Corner-Law model for scaled gap widths up to approximately $\tilde{w}_i\approx0.2$. Beyond this range, the contribution of the edge fields becomes more significant, and the predictions of the two models increasingly diverge. 

An additional advantage of the Cubic model is its simpler analytical form, which reduces the computational cost of evaluating the Laplace solution [Eq.~\eqref{eq:Laplace_Cubic}] compared with the Corner-Law model [Eq.~\eqref{eq:Laplace_Corner}], whose expression involves fractional-power terms. Thus, for scaled gap widths $\tilde{w}_i\lesssim0.2$, where the two models yield very similar predictions, the Cubic model offers a computationally efficient alternative to the Corner-Law model. For larger gap widths, or when a more accurate representation of the electric field near electrode edges is required, the Corner-Law model is preferable.
\begin{table}
\centering
    \caption{Optimized parameters of the sample trap with finite external boundaries, evaluated in the zero-gap limit and at finite gap widths using three different gap-potential models, corresponding to the optimized harmonic configuration defined by Eq.~\eqref{eq:Opt_path_1}.}
    \label{tab:optimized_parameters_various_gap_models}
\begin{tabular}{lcccc}
\hline\hline
\multicolumn{5}{c}{
\begin{tabular}{c}
Optimized for $V_0 = -0.3591\,\mathrm{V}$, $f_z = 4\,\mathrm{GHz}$, $\rho_1 = 10\,\mu\mathrm{m}$, \\
External boundaries $\{\rho_c,z_c\}=\{250,\,750\}\,\mu\mathrm{m}$ \\
Scaled radii $\tilde{\rho}_i = \{1.0,\, 5.0,\, \tilde{\rho}_3\}$, \\
Imposed constraints $C_1 = 0$, $C_2 = 1$, and \\
$C_3 = C_4 = C_6 = 0 \Rightarrow a_2 = a_3 = a_4 = a_5 = 0$.
\end{tabular}
}\\
\hline
Parameter & \hspace{5pt} Zero-Gap & \hspace{5pt} Linear & \hspace{5pt} Cubic & \hspace{5pt} Corner-Law\\
\hline
$\tilde{w}_i$ & 0.0 & 0.5 & 0.5 & 0.5\\
$\tilde{\rho}_3$ & 9.1888 & 8.3662 & 8.2179 & 8.1983\\
$\tilde{V}_1$ & -29.4017 & -26.6593 & -26.1674 & -26.1025\\
$\tilde{V}_2$ & -30.0639 & -27.3107 & -26.8162 & -26.7511\\
$\tilde{V}_3$ & -72.7628 & -74.7453 & -75.3809 & -75.4727\\
$\tilde{z}_0$ & 1.4342 & 1.3714 &  1.3592 & 1.3575\\
$a_6$ & -0.0039 & -0.0043 & -0.0044 & -0.0044\\
$\Delta f_z \,(\mathrm{Hz})$ & 0.656 & 0.731 & 0.750 & 0.752\\
$\sigma_f~(\mathrm{Hz})$ & 2.860 & $3.174$ & $3.257$ & $3.266$ \\
\hline\hline
\end{tabular}
\end{table}
\begin{figure}
    \centering
    \includegraphics[width=0.9\columnwidth]{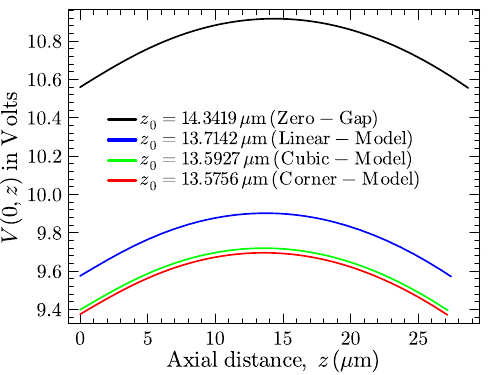}
    \caption{Axial potential of the sample trap along the $z$-axis for the zero-gap approximation and a finite gap width $w_i=5.0\,\mu\mathrm{m}$, evaluated using the different gap-potential models under fixed external boundaries $\{\rho_c,z_c\}=\{250,\,750\}\,\mu\mathrm{m}$. All cases satisfy Eq.~\eqref{eq:Opt_path_1} and exhibit near-harmonic behavior about their respective equilibrium positions $z_0$, with differences primarily reflected in the overall potential offset.}
    \label{fig:axial_pot_finite_gap_models_finite_bound}
\end{figure}

The thermal frequency spread $\sigma_f$ [Fig.~\ref{fig:Effect of finite gap widths with different gap models}(f)] increases only moderately with gap width. This behavior arises from changes in the unconstrained higher-order potential coefficients, particularly $C_5$ and $C_8$, which contribute to the amplitude coefficient $a_6$ and thereby introduce small corrections to both $\Delta f_z$ and $\sigma_f$. Although finite gaps substantially modify the optimized geometry and electrode potentials through gap-induced fringe fields, as illustrated by the axial potentials in Fig.~\ref{fig:axial_pot_finite_gap_models_finite_bound}, the full constraint set of Eq.~\eqref{eq:Opt_path_1} remains satisfied, ensuring that $a_2=a_3=a_4=a_5=0$. Consequently, the remaining changes in $\Delta f_z$ and $\sigma_f$ arise primarily from the higher-order anharmonic contribution associated with $a_6$. 

The three gap-potential models yield closely agreeing predictions for $\tilde{w}_i\lesssim0.1$, indicating that the optimized trap parameters and resulting frequency characteristics are relatively insensitive to model choice in this regime. Beyond $\tilde{w}_i\approx0.1$, their predictions gradually diverge. Table~\ref{tab:optimized_parameters_various_gap_models} compares the optimized parameters at $\tilde{w}_i=0.5$, where the differences are more evident. The optimized $\tilde{\rho}_3$ obtained using the Linear model differs by approximately $2\%$ from the Corner-Law result, whereas the Cubic model differs by only $0.24\%$. The Cubic model's optimized electrode potentials differ by less than $0.25\%$ from the Corner-Law results, while the differences in $\Delta f_z$ and $\sigma_f$ are approximately $0.27\%$ and $0.28\%$, respectively. While all three models satisfy the optimization constraints within their respective approximations, this does not imply equal accuracy in representing the actual finite-gap potential. In particular, satisfying the constraints with the Linear model does not ensure that it captures the gap-induced fringe fields. Thus, model choice affects the optimized configuration at larger gap widths, but the Cubic model closely approximates the Corner-Law model for the configurations considered here.

As demonstrated by comparison with the FEM-simulated gap potentials in Sec.~\ref{FEM_model}, the Corner-Law model provides the most accurate analytical representation among the three models considered at very small gap widths. The Cubic model also closely reproduces the Corner-Law potential profiles, even at $\tilde{w}_i=0.5$, making it a useful approximation. However, the agreement of both models with FEM deteriorates at sufficiently large gap widths. This emphasizes the importance of accurately representing the spatial potential variation across finite electrode gaps, particularly when designing traps with relatively large gap-to-electrode ratios.

\section{Comparison of Analytic Results with Finite-Element Simulations}
\label{simcomp}
Finite-element-method (FEM) simulations are performed using COMSOL Multiphysics\textsuperscript{\textregistered} \cite{comsol63} to assess the analytical treatment of finite electrode gaps and to benchmark the resulting trap optimization against the full three-dimensional electrostatic problem. In particular, the simulations address two questions: 
\begin{itemize}
    \item whether the analytical gap-potential approximation adequately describes the electrostatic boundary conditions of a realistic planar trap, and 
    \item whether the harmonicity obtained from the analytical optimization is preserved when the optimized design is implemented with finite electrode dimensions.
\end{itemize}

\subsection{General Implementation, Numerical Convergence, and Consistency}
\subsubsection{Implementation}
The three-dimensional planar-trap geometry is constructed in COMSOL using the optimized geometry and electrode potentials obtained from the analytical Corner-Law gap model (see Fig.~\ref{fig:Effect of finite gap widths with different gap models}), unless stated otherwise. Rotational symmetry is employed to reduce computational cost. Throughout this section, unless stated otherwise, the studies are performed using the analytically optimized parameters of a sample planar trap with external boundaries $\{\rho_c,z_c\}=\{250\,\mu\mathrm{m}, 750\,\mu\mathrm{m}\}$ and a gap width of $w_i=0.01\,\mu\mathrm{m}$ ($\tilde{w}_i=0.001$), which approaches the infinitesimal-gap limit. The corresponding optimized geometrical parameters and electrode potentials can be inferred from Fig.~\ref{fig:Effect of finite gap widths with different gap models}. These parameters are kept fixed and used in the subsequent studies to provide a consistent basis for comparison. Particular attention is given to resolving strong electric-field gradients near the electrode edges and around the equilibrium position $z_0$. The electrostatic solver tolerance is set to $10^{-12}$.


The simulated axial potential is transferred to MATLAB~\cite{MATLAB}, which controls COMSOL through Live-link, performs the optimization, and carries out the subsequent analysis. The potential is fitted to Eq.~\eqref{eq:axial_pot_Ck} over approximately $70\%$ of the region surrounding $z_0$, using potential coefficients up to order 20.

\subsubsection{Realism of Gap Treatment: Analytical vs. FEM Simulation}
\label{real}
In the analytical treatment, the finite electrode gaps are represented by an effective surface-gap model in which the electrodes are treated as infinitely thin conducting boundaries. This approximation assumes that the electrode height (gap depth) is sufficiently large that axial variations of the electric field within the gap can be neglected, and that the electrode widths are much larger than the gap width, so that the gap is locally bounded by semi-infinite electrodes. Under these conditions, the gap potential is approximated using the boundary conditions in Eq.~\eqref{eq19}, where the potential at the gap center is taken as the average of the potentials of the two adjacent electrodes. This greatly simplifies the electrostatic boundary-value problem and permits an analytical solution.

The actual three-dimensional geometry, in contrast, has finite electrode heights (corresponding to the gap depths), finite electrode widths, and coupled radial--axial boundary conditions, forming a mixed boundary-value problem that generally requires numerical treatment. The FEM simulations therefore provide a direct assessment of the analytical approximation by explicitly accounting for these finite geometric dimensions.

To quantify these effects, we define the axial and radial aspect ratios
\[
AR_z=\frac{\text{electrode height}}{\text{gap width}},
\qquad
AR_r=\frac{\text{electrode width}}{\text{gap width}}.
\]
For the $i^{\mathrm{th}}$ gap, which is bounded by inner and outer electrodes (Fig.~\ref{ith_gap}), the latter is characterized by $AR^i_{\mathrm{r,in}}$ and $AR^i_{\mathrm{r,out}}$.

Simulations with different electrode heights show that the axial potential becomes essentially independent of electrode height for $AR_z\gtrsim3$--$4$. We therefore use $AR_z=5$ throughout the FEM calculations, ensuring negligible finite-height effects. The radial aspect ratios have a more pronounced influence. When both $AR^i_{\mathrm{r,in}}$ and $AR^i_{\mathrm{r,out}}$ are large, the neighbouring electrodes closely approximate the semi-infinite boundaries assumed analytically. As these ratios decrease, finite electrode widths modify the local gap-potential distribution, with the effect becoming increasingly important for larger gap widths. This provides the basis for the direct analytical--FEM comparison as discussed in Sec.~\ref{FEM_model}.

\subsubsection{Convergence}
\begin{figure}[h]
\centering
\includegraphics[width=0.9\columnwidth]{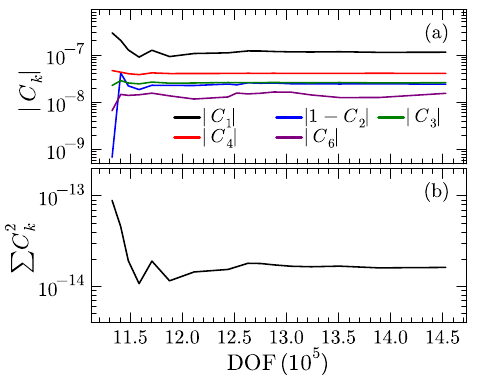}
    \caption{Mesh convergence of (a) the simulated potential coefficients $C_k\ \{C_1, 1-C_2, C_3, C_4, C_6\}$ and (b) their sum of squares as functions of DOF, obtained by progressively refining the mesh near the electrode corners and around the equilibrium position.}
\label{fig:mesh_conv}
\end{figure}

Reliable extraction of the potential coefficients $C_k$ requires sufficient mesh refinement near the electrode corners and around $z_0$. The FEM model uses $\mu\mathrm{m}$ as its global spatial unit. Fig.~\ref{fig:mesh_conv} shows the convergence of the coefficients $C_k$ and their sum of squares with increasing degrees of freedom (DOF). Although the coefficients initially exhibit fluctuations, they progressively stabilize with increasing DOF, as also reflected in their sum of squares. Based on this convergence study, similar mesh settings are adopted for subsequent simulations, with a maximum element size of $5\,\mu\mathrm{m}$ and a minimum element size of $10^{-5}\,\mu\mathrm{m}$ near the electrode corners and around the equilibrium position. Further refinement below this minimum size is avoided because it leads to unreliable meshing and limits the accuracy with which $z_0$ can be determined. These settings correspond to approximately $14.5\times10^5$ DOF.
\begin{figure}[b]
\centering
\includegraphics[width=0.9\columnwidth]{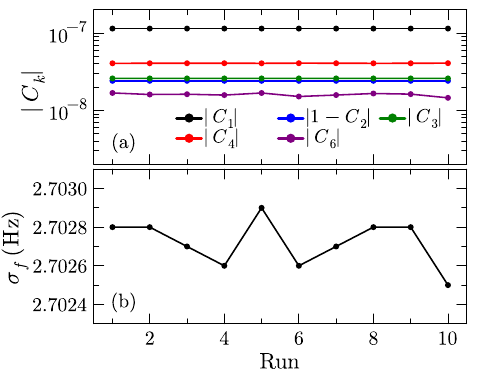}
   \caption{Repeatability of (a) the potential coefficients $C_k$ and (b) the resulting thermal frequency spread $\sigma_f$ at $4\,\mathrm{K}$, retaining terms up to $a_6$, over ten repeated simulations with identical input parameters.}
\label{fig:repeatability}
\end{figure}
\subsubsection{Repeatability}
In addition to mesh convergence, repeatability is essential for reliable FEM-based optimization, as numerical fluctuations can affect the optimizer's search path and lead to inconsistent solutions. We therefore adopt a cubic finite-element formulation, which provides more stable higher-order potential coefficients than the quadratic formulation, although both formulations converge. Fig.~\ref{fig:repeatability} illustrates the variations in the coefficients $C_k$ and thermal frequency spread $\sigma_f$ over ten independent COMSOL evaluations with identical geometry, electrode potentials, and mesh settings. The coefficients and frequency spread exhibit very small variations across these repeated simulation runs. In particular, the frequency spread has a standard deviation of the order of $10^{-4}$ Hz and a relative standard deviation of approximately $0.005\%$, while the relative standard deviations of the coefficients, except for $C_6$, remain below approximately $0.1\%$. Although $C_6$ exhibits a relatively large variation of about $4.5\%$, its absolute standard deviation remains of the order of $10^{-10}$. This sensitivity, greater than that of the lower-order coefficients, reflects the amplification of small run-to-run numerical fluctuations in the FEM-computed potential during the extraction of higher-order coefficients. Despite this variation and the associated limitation in the accuracy of $z_0$, highly harmonic trap parameters can still be determined reliably, as the overall optimization results exhibit good repeatability. These results confirm the numerical repeatability and stability achieved with the adopted FEM formulation and mesh settings.

\subsubsection{Parameter Sensitivity}
Parameter sensitivity is important for determining whether the optimizer can resolve sufficiently small parameter adjustments to achieve the required trap harmonicity. Adjustments below the effective numerical resolution of the FEM model may remain undetected, preventing further reduction of the potential coefficients.

To distinguish parameter-induced changes from numerical fluctuations in COMSOL, each parameter, $\rho_3$ and $V_i$ ($i=1,2,3$), is perturbed individually over progressively increasing step sizes. The perturbation ranges are $10^{-10}$ to $10^{-2}\mu\mathrm{m}$ for $\rho_3$ and $10^{-10}$ to $10^{-2}\mathrm{V}$ for the electrode potentials. At each step, the objective function is evaluated and compared with its nominal mean, $\overline{OBJ}$, obtained from ten repeated simulations using identical, unperturbed parameters. The objective function and its absolute deviation from the nominal mean are defined as
\[
\begin{aligned}
OBJ &= C_1^2+(1-C_2)^2+C_3^2+C_4^2+C_6^2,\\
\Delta OBJ &= \left|OBJ-\overline{OBJ}\right|,
\end{aligned}
\]
where $\overline{OBJ}$ provides a reference for assessing the numerical fluctuations associated with COMSOL repeatability.

\begin{figure}[h]
\centering
\includegraphics[width=0.9\columnwidth]{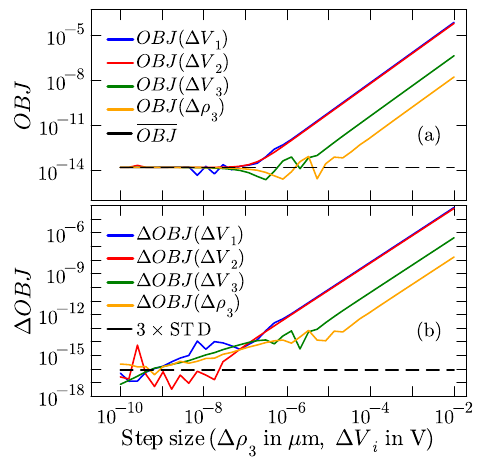}
    \caption{Sensitivity of the FEM model to perturbations in the optimized parameters: (a) objective function, \(OBJ\), and (b) its absolute deviation from the nominal mean, \(\Delta OBJ\), as functions of the perturbation step size for \(V_1\), \(V_2\), \(V_3\), and \(\rho_3\). The dashed lines indicate the nominal mean objective, \(\overline{OBJ}\), in (a) and the detection threshold, \(3\times STD\), in (b). Both quantities are calculated from ten evaluations using identical, unperturbed parameters.}
\label{fig:sensitivity}
\end{figure}

As shown in Fig.~\ref{fig:sensitivity}, the nominal objective is approximately $1.63\times10^{-14}$, with a standard deviation of $STD=2.90\times10^{-17}$. Accordingly, a detection threshold of $3\times STD\approx8.69\times10^{-17}$ is adopted to distinguish parameter-induced changes from numerical fluctuations. Perturbations producing $\Delta OBJ$ below this threshold cannot be reliably distinguished from the COMSOL noise floor.

For small step sizes ($10^{-10}$ to $10^{-8}$), the objective remains close to its nominal value, while $\Delta OBJ$ exhibits irregular fluctuations without a clear systematic trend. These variations generally remain below the detection threshold, indicating that such perturbations cannot be reliably distinguished from numerical noise. With increasing step size, the objective begins to depart from its nominal value, although some non-monotonic variations persist at intermediate steps. At larger step sizes, $\Delta OBJ$ exceeds the threshold and exhibits a clear, systematic increase. The approximate minimum perturbations required to produce a detectable response are $10^{-7}\,\mathrm{V}$ for $V_1$ and $V_2$, $10^{-6}\,\mathrm{V}$ for $V_3$, and $10^{-5}\,\mu\mathrm{m}$ for $\rho_3$. The greater sensitivity to $V_1$ and $V_2$ is attributed to their electrodes being closer to the trap symmetry axis ($z$-axis), where changes in electrode potential have a stronger influence on the axial potential distribution.

These results establish the practical sensitivity limits of the FEM model and, together with the repeatability analysis, provide a basis for assessing the reliability of the FEM-based optimization. While the analytical optimization satisfies the constraints of Eq.~\eqref{eq:Opt_path_1} to within $10^{-12}$, the FEM-based optimization is limited by numerical fluctuations and mesh discretization. Nevertheless, the potential coefficients can be reduced below the order of $10^{-8}$. The adopted mesh settings therefore provide adequate numerical repeatability and parameter sensitivity for optimizing highly harmonic traps and are retained for the subsequent FEM simulations.

\subsubsection{FEM-Based Optimization}
\label{optimizer}
Although MATLAB provides several built-in optimization algorithms, they were found unsuitable for achieving the required trap harmonicity because they offer insufficient control over Jacobian evaluation and parameter updates as the objective function approaches very small values. In this regime, numerical fluctuations in the FEM calculations can hinder reliable convergence. Therefore, a custom Gauss--Newton/Levenberg--Marquardt (GN--LM) algorithm was used, following the nonlinear least-squares framework described in Ref.~\cite{nocedal2006numerical}. It incorporates adaptive damping and a trust-region constraint to regulate parameter updates. The objective functions for the two FEM-based approaches are defined as follows. For FEM$(\rho_3+V)$-based optimization,
\[
OBJ(\mathbf{x})=\|\mathbf{F}(\mathbf{x})\|^2
=C_1^2+(1-C_2)^2+C_3^2+C_4^2+C_6^2,
\]
whereas, for FEM$(V)$-based optimization,
\[
OBJ(\mathbf{x})=\|\mathbf{F}(\mathbf{x})\|^2
=C_1^2+(1-C_2)^2+C_3^2+C_4^2.
\]
Both approaches are discussed in Sec.~\ref{constraint}. The Jacobian is evaluated using central finite differences, and the parameter correction is obtained from
\[
(J^TJ+\lambda D)\Delta\mathbf{x}=-J^T\mathbf{F},
\]
where $J$ is the Jacobian, $\lambda$ is the adaptive damping parameter, and $D$ is a diagonal scaling matrix constructed from the diagonal elements of $J^TJ$. Parameter updates are controlled by a trust-region constraint and a backtracking line search \cite{nocedal2006numerical}. Following an accepted step, the Jacobian is updated using Broyden's method, provided that the prescribed conditions on the objective values, parameter-step magnitude, and model accuracy are satisfied \cite{broyden1965class}. To account for FEM numerical fluctuations, a trial step is accepted only if the reduction in the objective exceeds the COMSOL repeatability threshold ($3\times\mathrm{STD}$), determined from repeated evaluations with identical, unperturbed parameters.

The stopping criterion for the optimizer is set to $OBJ<10^{-16}$, corresponding to individual residuals below $10^{-8}$ for the constraints in Eq.~\eqref{eq:Opt_path_1}, namely, $C_1$, $1-C_2$, $C_3$, $C_4$, and $C_6$. In the optimized cases considered here, these residuals typically reach values below $10^{-9}$, indicating that the constraints are satisfied to high numerical accuracy. Although the analytical optimization achieves residuals of the order of $10^{-12}$ or smaller, comparable accuracy is difficult to attain with FEM-based optimization because of numerical fluctuations and limitations imposed by the minimum achievable mesh element size. Nevertheless, the accuracy achieved with FEM is sufficient to recover the trap harmonicity degraded by finite gap widths.

For a scaled gap width of $\tilde{w}_i=0.001$, the initial FEM evaluation yields $C_k$ coefficients of the order of $10^{-7}$--$10^{-8}$ and a corresponding objective function (OBJ) of the order of $10^{-14}$ (see Figs.~\ref{fig:mesh_conv}, \ref{fig:repeatability}, and \ref{fig:sensitivity}). The FEM$(\rho_3+V)$-based optimization reduces the OBJ to below $10^{-16}$, with the individual $C_k$ coefficients reaching values below $10^{-9}$. The resulting optimized parameters nevertheless exhibit small deviations from their analytical values, of approximately $-10^{-4}$~$\mu$m for $\rho_3$, $-10^{-5}$~V for $V_1$, $-10^{-5}$~V for $V_2$, and $+10^{-6}$~V for $V_3$. This convergence is achieved in approximately 50 iterations, with a total computational time of about one hour. The initial $C_k$ coefficients can also be reduced to comparable levels through FEM$(V)$-based optimization.

For larger gap widths, however, the differences between the analytically and numerically optimized parameters become more pronounced, as discussed in Sec.~\ref{constraint}. At a scaled gap width of $\tilde{w}_i=0.5$, the optimizer requires more than 500 iterations to reduce the OBJ from an initial value of the order of $10^{-5}$ to below $10^{-16}$, with a total computational time exceeding 8 hours.

\subsection{Benchmarking of Analytical Results by FEM Simulation}
\label{FEM_model}
\begin{figure}[h]
    \centering
    \includegraphics[width=0.9\columnwidth]{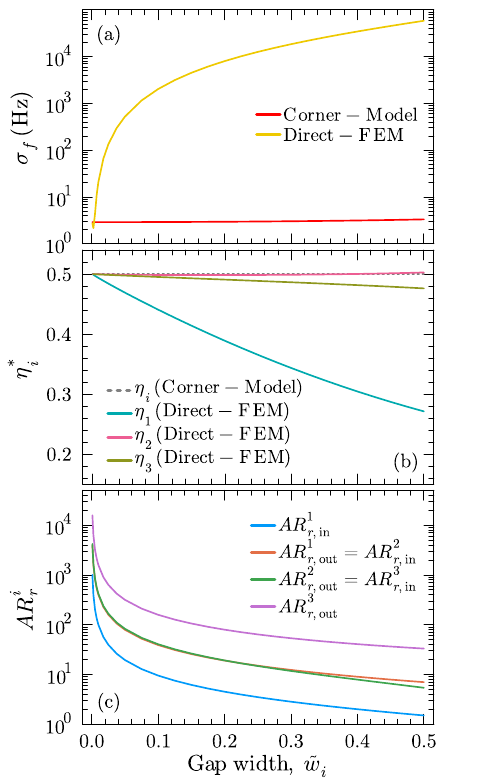}
    \caption{Comparison of (a) the thermal axial-frequency spread at $4\,\mathrm{K}$ predicted by the Corner-Law model and FEM, (b) the normalized average-potential location, $\eta_i^{*}=\tilde{\xi}_i/\tilde{w}_i$, and (c) the radial aspect ratios of the three gaps ($AR_z=5$), as functions of the scaled gap width using the analytically optimized Corner-Law parameters from Fig.~\ref{fig:Effect of finite gap widths with different gap models}. Excellent agreement is obtained for small gap widths. With increasing gap width, the FEM frequency spread increasingly deviates from the analytical prediction as $\eta_1^{*}$ shifts away from the midpoint, coinciding with a decrease in the radial aspect ratio of the first gap.}
    \label{fig:comp_sprad_eta_AR_vs_gap}
\end{figure}

To benchmark the analytical optimization, the optimized electrode geometry and voltages obtained from the Corner-Law model are directly implemented in the FEM simulations without imposing the constraints of Eq.~\eqref{eq:Opt_path_1}. The simulated axial potential is fitted using Eq.~\eqref{eq:axial_pot_Ck}, and the resulting coefficients $C_k$ are used to calculate the thermal axial-frequency shift and spread at $4\,\mathrm{K}$ from Eqs.~\eqref{eq:thermal_shift} and \eqref{eq:thermal_spread}, retaining amplitude-dependent terms through $a_6$. The comparison is performed over the scaled gap-width range $0.001\leq\tilde{w}_i\leq0.5$. Fig.~\ref{fig:comp_sprad_eta_AR_vs_gap} compares the Direct-FEM and analytical thermal frequency spreads, together with the normalized average-potential location $\eta_i^{*}=\tilde{\xi}_i/\tilde{w}_i$ and the radial aspect ratios of the three gaps.

As shown in Fig.~\ref{fig:comp_sprad_eta_AR_vs_gap}(a), the analytical and FEM results agree extremely well for small gap widths, confirming the validity of the analytical treatment in this limit. With increasing gap width, however, the discrepancy grows progressively. The origin of this deviation is evident from Fig.~\ref{fig:comp_sprad_eta_AR_vs_gap}(b), which shows the normalized average-potential location $\eta_i^{*}=\tilde{\xi}_i/\tilde{w}_i$, where $\tilde{\xi}_i$ is measured from the inner edge of the $i$th gap to the point at which the potential equals $(V_i+V_{i+1})/2$. The analytical model assumes $\eta_i^{*}=0.5$ (Eq.~\eqref{eq19}), corresponding to the midpoint of each gap. For small gap widths, all three gaps satisfy $\eta_i^{*}\approx0.5$. 

As the gap width increases, however, $\eta_1^{*}$ shifts progressively toward the inner electrode, while $\eta_2^{*}$ and $\eta_3^{*}$ remain close to the midpoint. This behavior is associated with the comparatively small radial aspect ratio of the first gap, as shown in Fig.~\ref{fig:comp_sprad_eta_AR_vs_gap}(c). Because the first gap is located at the smallest radius ($\tilde{\rho}_1=1.0$), its radial aspect ratio decreases most rapidly with increasing gap width, making the midpoint approximation progressively less accurate. Consequently, the FEM axial potential no longer satisfies the constraints of Eq.~\eqref{eq:Opt_path_1}; finite values of $C_3$, $C_4$, and $C_6$ remain, resulting in a larger amplitude dependence of the axial frequency than predicted analytically.

\begin{figure}[h]
    \centering
    \includegraphics[width=0.9\columnwidth]{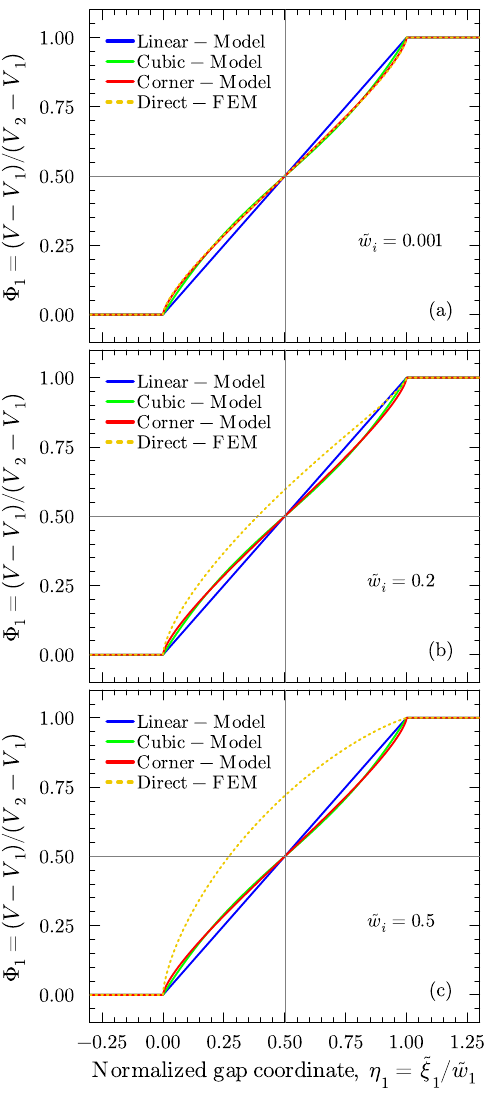}
    \caption{Comparison of the first-gap potential predicted by the Linear, Cubic, and Corner-Law models with FEM for (a) $\tilde{w}_i=0.001$, (b) $\tilde{w}_i=0.2$, and (c) $\tilde{w}_i=0.5$. For $\tilde{w}_i=0.001$, the Cubic and Corner-Law models closely reproduce the FEM profile, with $\eta_1^{*}\approx0.5$. For $\tilde{w}_i=0.2$ and $\tilde{w}_i=0.5$, the FEM profile becomes asymmetric and $\eta_1^{*}$ shifts toward the inner electrode, whereas the analytical models retain the midpoint condition of Eq.~\eqref{eq19}, leading to the observed discrepancy between the FEM and analytical models.}
    \label{fig:gap_pot_distribution_for_gap1}
\end{figure}

This effect is illustrated more directly in Fig.~\ref{fig:gap_pot_distribution_for_gap1}, which compares the potential distribution across the first gap for $\tilde{w}_1=0.001$, $0.2$ and $0.5$, corresponding to physical gap widths of $0.01\,\mu\mathrm{m}$, $2\,\mu\mathrm{m}$ and $5\,\mu\mathrm{m}$ for $\rho_1=10\,\mu\mathrm{m}$. For $\tilde{w}_1=0.001$, the Cubic and Corner-Law models closely reproduce the FEM gap potential, with $\eta_1^{*}\approx0.5$, and the corresponding thermal frequency spreads are nearly identical at approximately $3\,\mathrm{Hz}$. In contrast, for $\tilde{w}_1=0.2$, and $0.5$, the FEM gap potential becomes noticeably asymmetric and the average-potential point shifts toward the inner electrode. This departure from the midpoint condition produces finite anharmonic coefficients in the fitted FEM potential and increases the thermal frequency spread to approximately $8\,\mathrm{kHz}$, and $58\,\mathrm{kHz}$.

\begin{figure}[h]
    \centering
    \includegraphics[width=0.9\columnwidth]{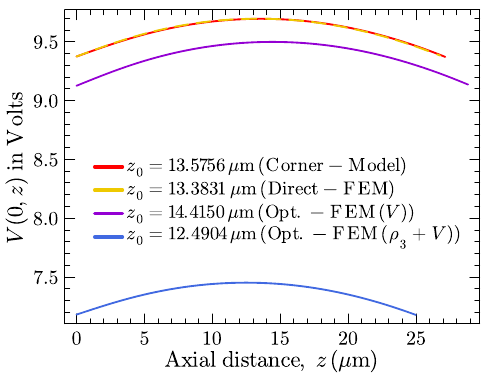}
    \caption{Comparison of the axial potentials of the sample trap along the $z$-axis for the scaled gap width $\tilde{w}_i=0.5$ obtained using the Analytical Corner-Model, FEM, FEM(V), and FEM($\rho_3+V$). The Direct-FEM result directly implements the analytically optimized geometry and electrode voltages. FEM(V) re-optimizes only the electrode voltages with the geometry fixed, whereas FEM($\rho_3+V$) simultaneously re-optimizes the electrode radius $\rho_3$ and the electrode voltages to satisfy Eq.~\eqref{eq:Opt_path_1}. The corresponding trap parameters are listed in Table~\ref{tab:analytical_sim_opt_sim_parameters}.}
    \label{fig:axial_pot_corner_sim_non_opt_sim_opt}
\end{figure}

\begin{figure}[h]
    \centering
    \includegraphics[width=0.9\columnwidth]{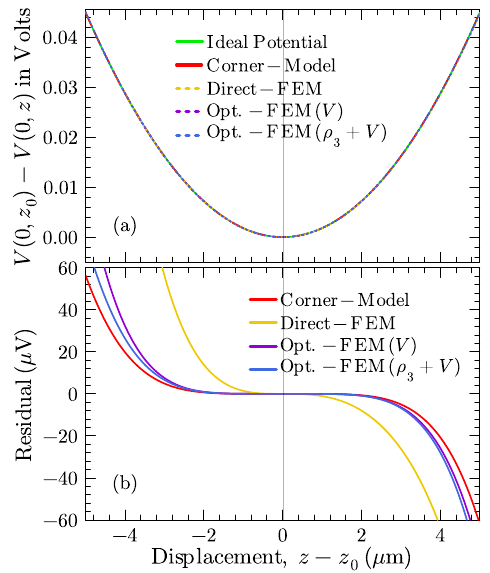}
    \caption{Comparison of the axial potentials of the sample trap for the scaled gap width $\tilde{w}_i=0.5$ obtained using the Analytical Corner-Model, FEM, FEM(V), and FEM($\rho_3+V$). (a) Axial potentials together with the corresponding ideal harmonic potential about $z_0$. (b) Deviations of the respective axial potentials from the ideal harmonic potential as functions of $(z-z_0)$. The corresponding trap parameters are listed in Table~\ref{tab:analytical_sim_opt_sim_parameters}.}
    \label{fig:axial_pot_res_corner_sim_opt_sim}
\end{figure}

The resulting effect on the axial trapping potential is shown in Fig.~\ref{fig:axial_pot_corner_sim_non_opt_sim_opt} for $\tilde{w}_i=0.5$. Although the Direct-FEM and analytical Corner-Law potentials appear nearly identical over the trapping region, their deviations from the corresponding ideal harmonic potentials, shown in Fig.~\ref{fig:axial_pot_res_corner_sim_opt_sim}, reveal a clear difference in higher-order harmonic content. The Direct-FEM potential departs increasingly from harmonicity away from the equilibrium position $z_0$, whereas the analytical potential remains nearly harmonic. This difference originates from the finite $C_3$, $C_4$, and $C_6$ coefficients generated by the breakdown of the midpoint approximation in the first gap.

A quantitative comparison for $\tilde{w}_i=0.5$ ($w_i=5\,\mu\mathrm{m}$) is given in Table~\ref{tab:analytical_sim_opt_sim_parameters}. The analytically optimized coefficients satisfy Eq.~\eqref{eq:Opt_path_1} to within $10^{-12}$, whereas the Direct-FEM coefficients remain finite, typically in the range $10^{-5}$--$10^{-3}$. 
\begin{table}[t]
    \centering
    \caption{
    Comparison of the sample-trap parameters obtained using the Analytical Corner-Model, FEM, FEM(V), and FEM($\rho_3+V$) for the scaled gap width $\tilde{w}_i=0.5$. FEM directly implements the analytically optimized geometry and electrode voltages in the FEM model. FEM(V) re-optimizes only the electrode voltages, whereas FEM($\rho_3+V$) simultaneously re-optimizes the electrode radius ($\rho_3$) and voltages to satisfy Eq.~\eqref{eq:Opt_path_1}.
    }
    \label{tab:analytical_sim_opt_sim_parameters}
\begin{tabular}{lcccc}
\hline\hline
\multicolumn{5}{c}{
\begin{tabular}{c}
Optimized for $V_0=-0.3591\,\mathrm{V}$, $f_z=4\,\mathrm{GHz}$, $w_i=5\,\mu\mathrm{m}$,\\
External boundaries $\{\rho_c,z_c\}=\{250\,\mu\mathrm{m},\,750\,\mu\mathrm{m}\}$.
\end{tabular}
}\\
\hline
Parameter & Analytical & FEM & FEM(V) & FEM($\rho_3+V$) \\
\hline
$\rho_1$\,($\mu\mathrm{m}$) & 10.0 & 10.0 & 10.0 & 10.0\\
$\rho_2$\,($\mu\mathrm{m}$) & 50.0 & 50.0 & 50.0 & 50.0\\
$\rho_3$\,($\mu\mathrm{m}$) & 81.9829 & 81.9829 & 81.9829 & 63.9344 \\
$V_1$\,(V) & 9.3743 & 9.3743 & 9.1258 & 7.1808\\
$V_2$\,(V) & 9.6072 & 9.6072 & 9.3514 & 7.3746 \\
$V_3$\,(V) & 27.1049 & 27.1049 & 26.9864 & 38.2526 \\
$z_0$\,($\mu\mathrm{m}$) & 13.5756 & 13.3830 & 14.4150 & 12.4904\\
$a_2$ & $0.0000$ & $0.0075$ & $\sim10^{-9}$ & $\sim10^{-10}$ \\
$a_3$ & $0.0000$ & $\sim10^{-5}$ & $0.0000$ & $0.0000$ \\
$a_4$ & $0.0000$ & $0.0020$ & $0.0045$ & $\sim10^{-9}$\\
$a_5$ & $0.0000$ & $-0.0001$ & $\sim10^{-11}$ & $0.0000$ \\
$a_6$ & $-0.0044$ & $-0.0058$ & $-0.0032$ & $-0.0079$ \\
$\Delta f_z$\,(Hz) & $\sim0.7$ & $\sim58000$ & $\sim133$ &$\sim1.3$ \\
$\sigma_f$\,(Hz) & $\sim3.2$ & $\sim58000$ & $\sim298$ &$\sim5.9$ \\
\hline\hline
\end{tabular}
\end{table}

\begin{figure}[h]
    \centering
    \includegraphics[width=0.9\columnwidth]{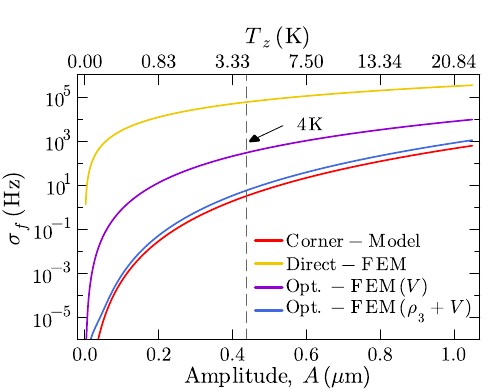}
    \caption{Comparison of the thermal axial-frequency spread, $\sigma_f$, as a function of oscillation amplitude (bottom axis) and the corresponding axial temperature $T_z$ (top axis) for the Analytical Corner-Model, Direct-FEM, Opt.-FEM(V), and Opt.-FEM($\rho_3+V$) at the scaled gap width $\tilde{w}_i=0.5$. The dashed line marks the thermal oscillation amplitude corresponding to $T_z=4\,\mathrm{K}$. The corresponding trap parameters are listed in Table~\ref{tab:analytical_sim_opt_sim_parameters}.}
    \label{fig:thermal_freq_spreads_corner_sim_opt_sim}
\end{figure}

The residual anharmonicity in the Direct-FEM potential produces a pronounced amplitude dependence of the axial frequency. At $4\,\mathrm{K}$, the absolute thermal frequency shift increases from approximately $0.7\,\mathrm{Hz}$ for the analytical model to $\sim58\,\mathrm{kHz}$ for Direct-FEM, while the thermal frequency spread increases from $\sim3.2\,\mathrm{Hz}$ to $\sim58\,\mathrm{kHz}$. Although the absolute difference is large, the corresponding frequency variation remains only $\sim10^{-5}$ relative to the $4\,\mathrm{GHz}$ axial frequency. The temperature dependence of $\sigma_f$ is shown in Fig.~\ref{fig:thermal_freq_spreads_corner_sim_opt_sim}; the Direct-FEM result remains above the analytical prediction over the full temperature range because of the residual anharmonicity of the simulated potential.

The comparison above shows that directly transferring the analytically optimized design to the FEM geometry does not preserve the harmonic constraints of Eq.~\eqref{eq:Opt_path_1}, particularly at large gap widths. This motivates a constraint-based FEM re-optimization using the custom GN--LM algorithm described in Sec.~\ref{optimizer} to compensate for deviations introduced by the finite three-dimensional geometry.
Two different re-optimization strategies are considered: 
\begin{itemize}
    \item {Only the electrode voltages are adjusted while the electrode geometry is kept fixed (``{in-operando tuning}''), we will label this ``FEM(V)''. This represents the practically relevant case in which the trap has already been fabricated and only the operating voltages can be modified. This approach tests the extent to which the residual anharmonicity can be compensated electrically. With the geometry fixed, voltage tuning can impose the lower-order conditions $C_1=0$, $C_2=1$, $C_3=0$, and $C_4=0$, thereby suppressing $a_2$, $a_3$, and $a_5$, while $C_6$ remains unconstrained and $a_4$ generally remains finite.}
\item{The electrode voltages and the third electrode radius $\rho_3$ are optimized simultaneously (``{ex-operando tuning}''), we will label this ``FEM($\rho_3+V$)''. The additional geometrical degree of freedom allows the full harmonic configuration of Eq.~\eqref{eq:Opt_path_1}, including $C_6=0$, to be imposed directly on the FEM solution. }
\end{itemize}

The implementation and resulting performance of these two re-optimization strategies are presented in the following section.

\subsection{Constraint-Based Re-Optimization by FEM Simulation}
\label{constraint}

To restore the optimized harmonic configuration defined by Eq.~\eqref{eq:Opt_path_1} in the FEM model and thereby suppress the residual anharmonicity, two constraint-based re-optimization approaches are investigated. The resulting optimized parameters are listed in Table~\ref{tab:analytical_sim_opt_sim_parameters} under the labels FEM(V) and FEM($\rho_3+V$).

In the first approach (``{in-operando tuning}''), only the three electrode voltages ($V_1$, $V_2$, and $V_3$) are treated as optimization variables, while all geometrical parameters are fixed at their analytically optimized values. The three voltage degrees of freedom allow four of the five constraints defining Eq.~\eqref{eq:Opt_path_1} to be imposed through the conditions
$C_1=0$, $C_2=1$, $C_3=0$, and $C_4=0$. These conditions yield $a_2=a_3=a_5=0$, while $C_6$ remains unconstrained and therefore $a_4$ generally remains finite.

In the second approach (``{ex-operando tuning}''), the third electrode radius $\rho_3$ is optimized simultaneously with the three electrode voltages, providing an additional geometrical degree of freedom. The resulting four adjustable parameters allow all five constraints of Eq.~\eqref{eq:Opt_path_1}, namely $C_1=0$, $C_2=1$, $C_3=0$, $C_4=0$, and $C_6=0$, to be imposed simultaneously. Consequently, $a_2=a_3=a_4=a_5=0$. The resulting axial potentials are compared with those of the Analytical Corner-Model and the Direct-FEM implementation in Figs.~\ref{fig:axial_pot_corner_sim_non_opt_sim_opt} and~\ref{fig:axial_pot_res_corner_sim_opt_sim}, while the corresponding trap parameters are presented in Table~\ref{tab:analytical_sim_opt_sim_parameters}.

The two re-optimized potentials have nearly identical local curvature around their respective equilibrium positions, although they differ in overall potential offset and in the location of the equilibrium position. Their residual potentials, however, reveal distinct higher-order harmonic content:

{In-operando tuning}, i.e., FEM(V), substantially suppresses the dominant lower-order anharmonicity by restoring $C_1=0$, $C_2=1$, $C_3=0$, and $C_4=0$. As a result, the absolute thermal axial-frequency shift at $4\,\mathrm{K}$ is reduced from approximately $58\,\mathrm{kHz}$ for the Direct-FEM solution to $133\,\mathrm{Hz}$. Because the geometry remains fixed, however, $C_6$ cannot be independently controlled and $a_4$ remains finite. Consequently, residual higher-order deviations persist away from the trap center, giving a thermal frequency spread of approximately $298\,\mathrm{Hz}$, nearly two orders of magnitude larger than the analytical value.

{Ex-operando tuning}, i.e., FEM($\rho_3+V$), satisfies all five constraints of Eq.~\eqref{eq:Opt_path_1}, reducing $a_2$, $a_3$, $a_4$, and $a_5$ to numerical precision. The resulting residual potential therefore closely follows that of the Analytical Corner-Model, with only a small deviation at larger oscillation amplitudes, see Fig.~\ref{fig:axial_pot_res_corner_sim_opt_sim}. This remaining difference arises because Eq.~\eqref{eq:Opt_path_1} does not constrain $C_5$ and $C_8$; their different values in the FEM solution lead to a different $a_6$. Consequently, the absolute thermal axial-frequency shift and spread at $4\,\mathrm{K}$ are reduced to approximately $1.3\,\mathrm{Hz}$ and $5.9\,\mathrm{Hz}$, respectively, compared with $0.7\,\mathrm{Hz}$ and $3.2\,\mathrm{Hz}$ for the analytical solution. 

As shown in Fig.~\ref{fig:thermal_freq_spreads_corner_sim_opt_sim}, the FEM($\rho_3+V$) and analytical curves remain nearly parallel over the full amplitude (temperature) range, differing mainly by a small, nearly constant offset. The larger magnitude of $a_6$ in the FEM($\rho_3+V$) solution ($-0.0079$ versus $-0.0044$ for the Analytical Corner-Model) accounts for this residual difference (see Table~\ref{tab:analytical_sim_opt_sim_parameters}). If terms only through $a_5$ are retained in the calculation of the thermal frequency shift and spread, the two predictions become identical within numerical precision. 
\begin{figure}[!h]
    \centering
    \includegraphics[width=0.9\columnwidth]{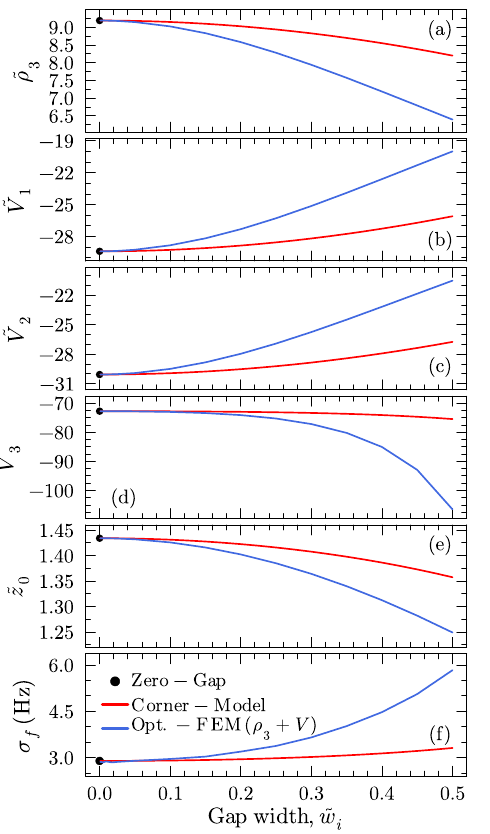}
    \caption{Comparison of the optimized parameters of the sample trap for fixed external boundaries $\{\rho_c,z_c\}=\{250,\,750\}\,\mu\mathrm{m}$
    as a function of scaled gap width $\tilde{w}_i$ obtained using the Analytical Corner-Model, and FEM($\rho_3+V$) under the constraints of Eq.~\eqref{eq:Opt_path_1}: (a) optimized $\tilde{\rho}_3$; (b)--(d) corresponding scaled electrode potentials $\tilde{V}_1$, $\tilde{V}_2$, and $\tilde{V}_3$; (e) equilibrium position $\tilde{z}_0$; and (f) thermal frequency spread $\sigma_f$ at $T_z=4\,\mathrm{K}$, calculated by retaining anharmonic coefficients through $a_6$. The zero-gap results (solid circles at $\tilde{w}_i=0$) are shown for reference.}
    \label{fig:comparison_FEM_corner}
\end{figure}

Further, sample trap parameters are optimized using FEM$(\rho_3+V)$ for selected scaled gap widths in the range $0.001\leq\widetilde{w}_i\leq0.5$, with the analytically optimized parameters as initial guesses. Fig.~\ref{fig:comparison_FEM_corner} compares the resulting parameters with those from the Analytical Corner-Model. The two approaches yield identical parameters in the zero-gap limit, but their differences increase with gap width due to the breakdown of the midpoint approximation in the first gap, as discussed above. Consequently, increasingly larger adjustments to $\rho_3$ and the electrode voltages are required in the FEM optimization to satisfy the constraints of Eq.~\eqref{eq:Opt_path_1}. Although both solutions satisfy these constraints, $C_5$ and $C_8$ remain unconstrained, leading to different $a_6$ coefficients and thermal frequency spreads. The larger magnitude of $a_6$ in the FEM solution (see Table~\ref{tab:analytical_sim_opt_sim_parameters} for $\widetilde{w}_i=0.5$) accounts for its larger $\sigma_f$ and its more rapid increase for $\widetilde{w}_i\gtrsim0.1$ [Fig.~\ref{fig:comparison_FEM_corner}(f)]. Thus, the Analytical Corner-Model provides an excellent approximation for gap widths up to $\widetilde{w}_i<0.1$, whereas FEM-based re-optimization becomes increasingly important for accurately capturing finite-gap effects. 

These results demonstrate that in-operando voltage tuning can substantially suppress the lower-order anharmonicity but cannot, in general, eliminate the higher-order distortions introduced by the realistic three-dimensional geometry, because the electrode geometry is fixed after fabrication. This limitation becomes particularly important at large scaled gap widths, such as $\tilde{w}_i=0.5$. Thus, while the analytical optimization provides an excellent starting point for trap design, simultaneous FEM-based optimization of the electrode geometry and operating voltages is required to recover similarly harmonic performance under realistic finite-gap and finite-boundary conditions.

\section{Summary}
\label{summ}
Here, we briefly summarize the most important statements about the planar Penning trap optimization as detailed in the preceding sections.
\begin{itemize}
    \item Planar Penning traps by design lack axial symmetry, but are able to produce highly harmonic confining axial potentials around an equilibrium position above their surface (Secs.~\ref{one} and~\ref{two}).
    \item For the concentric trap geometry, as illustrated in Fig.~\ref{trap}, an analytic description of the resulting potential is possible. This requires appropriate treatment of electrode gaps and finite electrostatic boundaries, particularly for small traps where the relative gap widths are large.
    \item In contrast to common Penning traps, the depth $V_0$ of the axial confining potential is much smaller than the applied electrode voltages, usually by one to two orders of magnitude. The equilibrium position $z_0$,
    measured from the electrode plane, is commonly of the same order as the characteristic trap dimension $\rho_1$. These features allow the confining potential to be generated from relatively large electrode voltages while maintaining a shallow potential-energy minimum, which can be advantageous for trap operation due to the insensitivity to voltage fluctuations.
    \item The required harmonicity of the potential for QIP is reached by tuning all of the trap geometry and voltages to the point where axial oscillation frequency widths due to the thermal motion of the confined electron are smaller than the frequency width of the resonant electronic non-destructive detection, such that the electron and its state can be detected and resonantly manipulated (Sec.~\ref{reqopt}).
    \item Sufficiently harmonic potentials in the above sense require at least 4 concentric electrodes (three gaps), since 2- and 3- electrode traps cannot be tuned for harmonic confinement. With 4 electrodes, the potential can be tuned to 5th order of the expansion in Eq.~\eqref{eq:freq_shift}. 
    \item In the current case, tuning to 5th order reduces the relative frequency shift between $T=0$ and $T=4$\,K to a level of about $1.6 \times 10^{-10}$ (Sec.~\ref{infbound}), such that higher-order tuning is likely not to further improve the overall situation since at that level other disturbances come into play that mask such improved stability.
    \item Tuning for high harmonicity usually involves all geometry parameters and electrode voltages. This is different from orthogonal traps \cite{gab89} and requires a highly harmonic planar trap to be designed for a specific value of the axial oscillation frequency. In-operando tuning of voltages can generally preserve the axial frequency and low-order harmonicity, but is not necessarily sufficient to preserve high harmonicity. 
    \item Concerning harmonicity, finite electrostatic boundaries become indistinguishable from infinite ones already on size scales a few times the overall size of the trap (Sec.~\ref{finbound}), and the effect of finite boundaries is mainly a constant offset of the trap potentials. For a fixed trap geometry, this offset can be eliminated not only in the infinite-boundary limit but also at a specific finite boundary location. The existence of such finite-boundary roots provides an additional mechanism for suppressing amplitude-dependent frequency shifts.
    \item Finite-height effects of the electrodes become negligible for electrode heights $h$ equal or greater than about three to four times the gap width $w$ (Sec.~\ref{real}).
    \item The role of the gaps between electrodes with regard to the required trap tuning for harmonicity increases with their relative size more than linearly. The various analytical models for the gap potentials differ in their predictions (Secs.~\ref{Infinitesimal_Gaps} to~\ref{Cubic_Model}). The cubic and corner-law models are more realistic than the linear model and particularly of course the infinitesimal-gap model and agree largely with one another.
    \item For small gap widths ($\tilde{w}_i \lesssim 0.05$), the optimized trap parameters are only weakly sensitive to both the presence of finite gaps and the choice of gap potential model, whereas larger gap widths require a more realistic description of the gap field, for which the corner-law model provides the most physically accurate representation. 
    \item The analytical and Direct-FEM simulation results agree extremely well for very small gap widths, confirming the validity of the analytical optimization in this limit. With increasing gap width, however, the discrepancy grows progressively. This motivates a constraint-based re-optimization by FEM Simulation (Sec.~\ref{constraint}).
    \item Such a re-optimization can be done by tuning of voltages only (``in-operando'') or by tuning of voltages and the geometry (``ex-operando''), with the former case having limits that can be overcome in the latter case. It allows to recover the analytical harmonic performance and the corresponding thermal frequency shift and spread under realistic finite-gap and finite-boundary conditions (Sec.~\ref{constraint}).
    \item Once re-optimized, even the weaker in-operando tuning keeps the thermal axial frequency width $\sigma_f$ up to temperatures of around 20\,K under $10^{-6}$ of the axial frequency $f_z$, such that quality factors $Q$ for resonant electron detection with tuned circuits could be as high as $10^6$ (Sec.~\ref{constraint}). Hence, the condition $\sigma_f \ll \gamma_z$ (Sec.~\ref{reqopt}) is fulfilled for all realistic choices of $Q$.
\end{itemize}
We note that this discussion has been focused on the axial potential only. Under non-ideal confinement conditions, effects other than anharmonicities of the electrostatic trap potential can lead to unwanted shifts of the axial oscillation frequency. The main culprits are drifts of the electrode voltages, geometrical tolerances, a spurious ellipticity of the effective potential around the trap center, and a misalignment of the electrode plane with respect to the magnetic field (a tilt). None of these are fundamental and can be minimized by technical efforts. Ellipticity is usually deemed negligible due to the high degree of obtainable cylindrical symmetry in trap manufacturing. Residual tilts in realistic setups are commonly of the order of a tenth of a degree and tend to slightly shift the axial frequency. Both effects have been quantitatively discussed in \cite{book} and the referenced work therein.

\section{Conclusion}
We have presented a careful analysis of the conditions required for harmonic confinement of single electrons in planar Penning traps that consist of concentric electrodes, with particular focus on cases where the gaps separating the electrodes are of non-negligible size. This is relevant especially for small traps below the mm size scale as proposed for the implementation of QIP in arrays of planar traps that store one electron each. A summary of the main results is given in Sec.~\ref{summ}.

It turns out that while planar Penning traps by design break the axial mirror symmetry of the electric field that ensures axial confinement, the region around the electrostatic potential around the equilibrium position can be made highly harmonic by appropriate choice of electrode sizes and voltages. Since planar traps are non-orthogonal by design, tuning of the trap generally involves all electrode sizes and voltages. This also means that an optimized harmonic trap of given size is fixed to one axial oscillation frequency of a stored particle.

Such planar traps feature an equilibrium position for particle storage on the central symmetry axis above the electrode plane. For electron confinement, this position corresponds to a maximum of the electrostatic potential and hence to a minimum of the electron potential energy. The corresponding confining potential depth is commonly one to two orders of magnitude smaller than the voltages applied to the individual electrodes. The distance of the equilibrium position from the electrode plane is usually comparable to the characteristic trap size, i.e., the diameter of the central disc electrode.

For the axial oscillation frequency distribution of the stored particle at non-zero temperature to be sufficiently small for the envisaged applications, a trap with four concentric electrodes (``three-gap trap'') is both required and sufficient, even for non-negligible gap sizes. Smaller numbers of electrodes are insufficient, higher numbers of electrodes increase the options for trap tunability but also the trap complexity.

With increasing relative gap size, the differences between different analytical models that describe the electrostatic potential in and due to the gaps, and between analytical models and finite-element simulations increase, to the point where even sophisticated analytical descriptions require a re-optimization by simulations. However, even for gap sizes that are far from negligible, the traps can be optimized by appropriate choice of sizes and voltages of the electrodes, albeit not by analytical means alone. 

Once a trap design is re-optimized by non-analytic FEM simulations as discussed, even the weaker in-operando tuning keeps the thermal axial frequency width $\sigma_f$ up to temperatures of around 20\,K under $10^{-6}$ of the axial frequency $f_z$, such that quality factors $Q$ for resonant electron detection with tuned circuits could be as high as $10^6$. Hence, the condition $\sigma_f \ll \gamma_z$ is fulfilled for all currently realistic choices of $Q$.

\bibliography{apssamp}

\end{document}